\documentclass[
reprint,
amsmath,amssymb,
aps,
nofootinbib,
]{revtex4-2}

\usepackage[T1]{fontenc}
\usepackage{graphicx}
\usepackage{mathtools}
\usepackage{amssymb}
\usepackage{amsthm}
\usepackage{xcolor}
\usepackage{nameref}
\usepackage[linkcolor=cyan,colorlinks=true,linktocpage=true]{hyperref}
\usepackage{physics}
\usepackage{standalone}
\usepackage{dsfont}
\usepackage{listings}
\usepackage{subcaption}
\usepackage{geometry}   
\usepackage{tikz}
\usepackage{tikz-3dplot}
\usetikzlibrary{decorations.pathreplacing}
\usetikzlibrary{positioning}
\usetikzlibrary{patterns}
\usetikzlibrary{shapes.geometric}
\usetikzlibrary{decorations.pathmorphing}
\usepackage{pgfplots}
\pgfplotsset{compat=1.18}

\usepackage{xcolor}

\definecolor{MyBlue}{HTML}{3D507D} 
\definecolor{MyLightBlue}{HTML}{5c6d96} 
\definecolor{MyGold}{HTML}{C0AB6A} 
\definecolor{MyGreen}{HTML}{39806f}

\let\originaleqref\eqref
\renewcommand{\eqref}{\originaleqref}
\numberwithin{equation}{section}
\newcommand{\figref}{FIG. \ref}
\numberwithin{figure}{section}
\newcommand{\listref}{Listing \ref}
\newcommand{\appref}{Appendix \ref}

\definecolor{codegreen}{rgb}{0,0.6,0}
\definecolor{codegray}{rgb}{0.5,0.5,0.5}
\definecolor{codepurple}{rgb}{0.58,0,0.82}
\definecolor{backcolour}{RGB}{225,231,245}

\lstdefinestyle{mystyle}{
	backgroundcolor=\color{backcolour},   
	commentstyle=\color{codegreen},
	keywordstyle=\bfseries\color{magenta},
	numberstyle=\tiny\color{codegray},
	stringstyle=\color{codepurple},
	basicstyle=\ttfamily\footnotesize,
	breakatwhitespace=false,         
	breaklines=true,                 
	captionpos=b,                    
	keepspaces=true,                 
	numbers=left,                    
	numbersep=5pt,                  
	showspaces=false,                
	showstringspaces=false,
	showtabs=false,                  
	tabsize=2
}

\newtheorem{theorem}{Theorem}[section]
\newtheorem{corollary}{Corollary}[theorem]
\newtheorem{lemma}[theorem]{Lemma}
\newtheorem{remark}[theorem]{Remark}

\usepackage{ragged2e}
\usepackage[compatibility=false]{caption}
\usepackage{lipsum}

\begin{document}

\preprint{APS/123-QED}

\title{Quantum capacity analysis of finite-dimensional lossy channels}

\author{Sofia Cocciaretto}
\email{sofia.cocciaretto@sns.it}
\affiliation{
	Scuola Normale Superiore, I-56126 Pisa,~Italy
}
\author{Vittorio Giovannetti}
\email{\\vittorio.giovannetti@sns.it}
\affiliation{NEST, Scuola Normale Superiore and Istituto Nanoscienze-CNR, p.za Cavalieri 7, Pisa, Italy}
\date{\today}

\begin{abstract}
	We investigate multi-level amplitude damping (MAD) channels as models of energy decay in finite-dimensional quantum systems, focusing on their quantum capacity. We provide a structural characterization of these channels and derive complete conditions for degradability and antidegradability, identifying precisely when the quantum capacity vanishes.
	We introduce a method to compute the quantum capacity beyond the degradable regime by combining monotonicity properties and reductions to lower-dimensional channels. We show that channels with completely damped levels admit a dimensional reduction that preserves capacity, enabling explicit evaluation in otherwise inaccessible regions.
	Our results extend known capacities to new parameter regimes and suggest that, in all analyzed cases, the quantum capacity is given by a single-letter formula. We also propose a conjecture on optimal encoding, indicating that certain levels can be excluded without loss of capacity.
\end{abstract}

\maketitle

\section*{Introduction}
{{Quantum Communication Theory (QCT) expands on the work of Shannon \cite{shannon}, who laid the mathematical foundations for the sharing of \textit{classical information}, to also encompass \textit{quantum information}. Just as we are used to think of \textit{bits} when we think of classical messages, such as text messages, music, videos etc., QCT is traditionally focused on the sharing of \textit{qubits} between parties across long distances; the mathematical objects that describe the transmission of quantum states are called \textit{quantum channels}.
The attention of this work is devoted to a particular class of quantum channels, i.e. \textit{Multi-level Amplitude Damping channels}, or MAD channels, which describe an energy decay of a \textit{qudit}, i.e. the $d$-dimensional equivalent of a qubit. The relevance of this kind of channels lies in their wide range of applicability: in fact, energy decays are commonplace over long distance communications, as can be exemplified by a fiber optic cable which is not completely isolated and can therefore lose a photon to the external environment.
Furthermore, the focus on higher-dimensional channels, i.e. channels built upon qudit transmission, which replace the more traditional qubits, is justified by the fact that recent results highlight how these kinds of architectures may present advantages both in terms of computation \cite{exp2} and communication \cite{exp1}.

Existing literature treated MAD channels for $d=2$, in which case the channels are called Amplitude Damping Channels (ADC) \cite{adc} and are very well understood, and for $d=3$ \cite{mad3}.

This work is a systematic study of MAD channels of arbitrary dimensionality, where we derive general properties (see Section \ref{section:setting}) with a special focus on \textit{degradability} and \textit{antidegradability} (Subsection \ref{section:degradability}) along with the development of a technique to compute the \textit{quantum capacity} (i.e. the rate at which quantum states can be reliably transmitted through the channels) of MAD's, even under conditions that normally render the computation of that capacity impossible (Section \ref{section:procedure}), a generalization of a procedure introduced in \cite{mad3}. We use $4$-dimensional MAD's as a sandbox to exemplify these results (Section \ref{sec:example}).

We want to highlight the main findings of this work:
\begin{itemize}
	\item Theorem \ref{theorem1} characterizes all the conditions under which $d$-dimensional MAD's are \textit{useless} for communication purposes, a result found through semi-analytical methods that involve Semi-Definite Programming (SDP) \cite{SDP}.
	\item Theorem \ref{Teorema2} allows for a symplification the computation of the quantum capacity of MAD channels when some of the levels are \textit{completely damped}.
	\item In Subsection \ref{subsection:conjecture} we formulate a conjecture on the optimal encoding of MAD's when only some of the conditions for their "uselessness" are satisfied, putting it to the test in a region of the parameter space of $3$-dimensional MAD's where the quantum capacity was not known but where the conjecture allows for its computation.
\end{itemize}

We also note that the computations performed for the quantum capacity are also valid for the \textit{classical private capacity}, as already shown in \cite{mad3}.}}

\section{MAD channels}
\label{section:setting}
\begin{figure}[]
	\centering
	\scalebox{0.6}{
		\begin{tikzpicture}[scale=2.5]
	\draw[black, thick] (4,0) -- (0,0) node [left] {\small{$\ket{0}$}};
	\draw[black, thick] (4,0.5) -- (0,0.5) node [left] {\small{$\ket{1}$}};
	\draw[black, thick] (4,1) -- (0,1) node [left] {\small{$\ket{2}$}};
	\draw[black, thick] (4,2) -- (0,2) node [left] {\small{$\ket{d-2}$}};
	\draw[black, thick] (4,2.5) -- (0,2.5) node [left] {\small{$\ket{d-1}$}};
	\draw[black, dotted, thick] (2,1.3) -- (2,1.7) node [left] {};
	\draw[blue, thick, ->] (1,0.5) -- (1,0) node [above left] {\footnotesize{$\gamma_{10}$}};
	\draw[codepurple, thick, ->] (1.5,1) -- (1.5,0.5) node [above left] {\footnotesize{$\gamma_{21}$}};
	\draw[codepurple, thick, ->] (2,1) -- (2,0) node [above left] {\footnotesize{$\gamma_{20}$}};
	\draw[red, thick, ->] (2.5,2.5) -- (2.5,2) node [above left] {\footnotesize{$\gamma_{d-1,d-2}$}};
	\draw[red, thick, dotted] (3,1.3) -- (3,1.7) node [left] {\footnotesize{$\gamma_{d-1,1}$}};
	\draw[red, thick] (3,2.5) -- (3,1.7) node [left] {};
	\draw[red, thick, ->] (3,1.3) -- (3,0.5) node [left] {};
	\draw[red, thick, dotted] (3.5,1.7) -- (3.5,1.3) node [left] {\footnotesize{$\gamma_{d-1,0}$}};
	\draw[red, thick] (3.5,2.5) -- (3.5,1.7) node [left] {};
	\draw[red, thick, ->] (3.5,1.3) -- (3.5,0) node [left] {};
\end{tikzpicture}
	}
	\caption{MAD channels represent decay processes, where each level of a system has a fixed probability of decaying onto a lower level. Here, a schematic depiction of a $d$-dimensional MAD is reported.
		This graph can be read from left to right in "chronological" order, thanks to \eqref{equation:MADdecomposition1}, where decays depicted with the same color happen simultaneously.
		In this case, the blue decay happens before the purple decays which in turn happen before the red ones.
	}
	\label{fig:madgraph4}
\end{figure}
Multi-level Amplitude Damping (MAD) channels were introduced in~\cite{mad3} as higher-dimensional generalization of ADC channels~\cite{adc}. They~are Linear Completely Positive, Trace Preserving~(LCPTP) qudit-to-qudit mapping~\cite{holevo_book} describing an energy decay process, under the assumptions that the energy gaps between the levels of the qudit are all different, see \figref{fig:madgraph4}.
Given a $d$-dimensional Hilbert space $\mathcal{H}$, spanned by the computational basis $\left\{|i\rangle\right\}_{i=0,\cdots,d-1}$,  a MAD channel admits a minimal Kraus representation in terms of following operators:
\begin{gather}
	\label{equation:definition:MADkrausset}
	\begin{gathered}
		K_{ij} \coloneqq  \sqrt{\gamma _{ji}}\op{i}{j} \quad 0\leq i<j\leq d-1,\\
		K_{00} \coloneqq \sum _{j=0} ^{d-1} \sqrt{\gamma_{jj}}\op{j}{j}.
	\end{gathered}
\end{gather}
For fixed $j$ and $i< j$, the parameters $\gamma_{ji}$  describe the probabilities of decay from level $| j\rangle$ onto level $|i\rangle$, while $\gamma _{jj}$'s describe the probabilities that level $| j\rangle$ will not decay during the transformation. Accordingly, they satisfy the constraints:
\begin{equation}
	\left\{\begin{array}{l}
		0\leq \gamma_{ji}\leq 1 \quad\forall \,0\leq i \leq j \leq d-1,                                            \\\\
		\sum_{i=0} ^{j}\gamma_{ji}=1 \;  \Longrightarrow \;  \gamma_{jj}\coloneqq 1-\sum_{i=0} ^{j-1}\gamma_{ji} , \\\\
		\gamma_{ji} = 0 \quad \forall i>j,\end{array}
	\right.\label{equation:gammajiproperties}
\end{equation}
which  in particular impose
\begin{eqnarray} \gamma_{00}=1\;. \end{eqnarray}
These quantities, which we name as \textit{transition probabilities} in what follows, can be grouped into an lower-triangular, $d\times d$ \textit{transition matrix}, defined by:
\begin{equation}
	\label{equation:definition:transitionmatrix}
	\Gamma \coloneqq
	\left(\begin{array}{ccccc}
			1             & 0             & 0             & 0           & \cdots \\
			{\gamma_{10}} & \gamma_{11}   & 0             & 0           & \cdots \\
			{\gamma_{20}} & {\gamma_{21}} & {\gamma_{22}} & 0           & \cdots \\
			{\gamma_{30}} & {\gamma_{31}} & \gamma_{32}   & \gamma_{33}
			              & \cdots                                               \\
			\vdots        & \vdots        & \vdots        &
			              & \ddots\end{array} \right) \;,
\end{equation}
%
which can be expressed in
compact form as
\begin{eqnarray}
	\label{equation:definition:transitionmatrixold}
	\Gamma &=&
	\sum _{j=0} ^{d-1}\sum_{i=0}^{j}\gamma _{ji}\op{j}{i} \\ \nonumber
	&=&  \mathds{1}+ \sum _{j=1} ^{d-1}\sum_{i=0}^{j-1}\gamma _{ji}\op{j}{i}-\sum _{j=1} ^{d-1}\sum_{i=0}^{j-1}\gamma_{ji}\op{j}{j}\;.
\end{eqnarray}
The relation between  the set ${\cal M}_d$ formed by the transition matrices and MAD channels is one-to-one: in particular given
$\Gamma\in{\cal M}_d$ of elements $\gamma_{ji}$ we define
it's associated MAD channel $\Phi _\Gamma$ the LCPTP transformation, which given an input density matrix  $\rho \in\mathfrak{S}(\mathcal{H})$ of components $\rho_{ij} := \langle i| \rho |j\rangle$,  outputs the state
\begin{eqnarray}
	\label{equation:definition:madchannelkraus}
	\Phi _{\Gamma} (\rho) &=& K_{00}\rho K_{00} ^{\dagger} +\sum _{j=1} ^{d-1}\sum_{i=0}^{j-1}K_{ij}\rho K_{ij} ^\dagger \\
	&=& \sum _{j=0}^{d-1}\sum_{i=0}^{d-1}\sqrt{\gamma _{jj}\gamma_{ii}}\rho _{ji}\op{j}{i}\nonumber\\
	&&+\sum_{j=1}^{d-1}\sum_{i=0}^{j-1}\gamma _{ji} \rho _{jj}\op{i}{i},\label{equation:definition:madchannelexpanded}
\end{eqnarray}
or equivalently
\begin{eqnarray}
	\label{equation:definition:madchannelkraus1}
	\Phi _{\Gamma} (|j\rangle\langle i|) 	=\left\{
	\begin{array}{ll} \sqrt{\gamma _{jj}\gamma_{ii}} |j\rangle\langle i| & \mbox{for $j\neq i$}\;, \\ \\
             \sum_{i'=0}^{j}\gamma _{ji'}  |i'\rangle\langle i'|     & \mbox{for $j = i$}\;.\end{array} \right.
\end{eqnarray}
Notice in particular that
the identity channel $\text{Id}$ is a special instance of MAD channel, whose
transition matrix is represented by the identity matrix, i.e.
\begin{eqnarray} \label{Identitychannel}
	\text{Id} = \Phi_{\mathds{1}} \;.
\end{eqnarray}
Another special example of MAD channel is the fully damping transformation
which maps all input states into
the ground level $|0\rangle$,
\begin{equation}
	\label{equation:definition:FULLY}
	\Phi_{\Gamma_{\emptyset}} (|j\rangle\langle i|) 	= \delta_{ji} |0\rangle\langle 0|\,, \quad \forall i,j =0, \cdots, d-1\;,
\end{equation}
characterized by the transition matrix $\Gamma_{\emptyset}$ with
elements $\omega_{ji} = \delta_{j0}$,
\begin{equation}
	\label{equation:definition:transitionmatrixFULL}
	\Gamma_{\emptyset} \coloneqq
	\left(\begin{array}{ccccc}
			1 & 0      & 0 & 0 & \cdots \\
			1 & 0      & 0 & 0 & \cdots \\
			  & \cdots                  \\
			1 & 0      & 0 & 0 & \cdots \\
			1 & 0      & 0 & 0 & \cdots\end{array} \right) \;.
\end{equation}
Furthermore one can easily verify that the set
${\cal M}_d$ is closed under convex convolutions, i.e.
\begin{eqnarray}
	&&\forall \Gamma, \Gamma' \in {\cal M}_d, \forall p\in [0,1] \\ \nonumber
	&& \qquad  \Longrightarrow \quad \Gamma(p):=p \Gamma + (1-p)\Gamma'
	\in {\cal M}_d\;.
\end{eqnarray}
This property is not shared by the set of
MAD channels which is not  convex. In particular,
unless the convolution is a trivial one,
the MAD channel associated with
$\Gamma(p)$ correspond to the convex combination only  the MAD channels associated with
$\Gamma$ and $\Gamma'$ only up to a concatenation with a dephasing LCPT map~\cite{holevo_book},  i.e.
\begin{eqnarray}\label{convexityconnection}
	\Delta \circ \Phi_{\Gamma(p)} =
	p \Phi_{\Gamma} + (1-p )\Phi_{\Gamma'}
	\;,
\end{eqnarray}
(see Appendix~\ref{appendix:comprules} for details).

\subsection{Composition rules}
\label{setting:composition}
A useful property of MAD channels is the fact that they are closed under concatenation.
Specifically given $\Phi _{\Gamma^{\prime}}$ and $\Phi _{\Gamma^{\prime\prime}}$  MAD channels, the transformation	$\Phi _{\Gamma^{\prime\prime}}\circ\Phi _{\Gamma^{\prime}}$
is also a MAD channel with transition matrix $\Gamma$ given by the ordered product of ${\Gamma^{\prime}}$ and ${\Gamma^{\prime\prime}}$:
\begin{equation}
	\label{equation:MADcomposition}
	\left\{ \begin{array}{l}
		\Phi _\Gamma = \Phi _{\Gamma^{\prime\prime}}\circ\Phi _{\Gamma^{\prime}}\;, \\\\
		\Gamma = \Gamma^{\prime} \Gamma^{\prime\prime}.\end{array} \right.
\end{equation}
The  proof of this fact is rather technical and we postpone it  in~\appref{appendix:comprules}: here we observe that together with Eq.~(\ref{Identitychannel}), it
implies that the set of MAD channel and
the set ${\cal M}_d$
are both  (non-Abelian) semi-groups with zero element represented by $\Phi_{\Gamma_{\emptyset}}$ and $\Gamma_{\emptyset}$ respectively.
The presence of the identity elements $\mathds{1}$ and $\Phi _{\mathds{1}}$ promotes $\mathcal{M}_d$ and the set of MAD channels, respectively, to monoids.
Another important consequence of Eq.~\eqref{equation:MADcomposition}  is the possibility of
identifying  decomposition rules for MAD channels: in what follows, we report the most useful ones for our scopes.

\subsubsection{Separated decays from increasing levels}\label{sec:sepdecay}
For $k < d$ define $\Phi_{\Gamma_k}$ and $\Phi_{\Gamma^{(k)}}$ the MAD channels characterized by
transition matrices
\begin{gather}
	\label{equation:gammadeco1}
	\begin{gathered}
		\Gamma _k \coloneqq \mathds{1}+ \sum_{i=0}^{k-1}\gamma _{ki}\op{k}{i} - \sum_{i=0}^{k-1}\gamma _{ki}\op{k}{k},\\
		\Gamma ^{(k)}\coloneqq \mathds{1}+\sum _{j=1} ^{k-1} \sum _{i=0} ^{j-1}\gamma _{ji}\op{j}{i}- \sum _{j=1} ^{k-1}\sum_{i=0}^{j-1} \gamma _{ji}\op{j}{j}.
	\end{gathered}
\end{gather}
The channel $\Phi_{\Gamma_k}$ represents a LPCTP transformation where only the level $\ket{k}$ is allowed to decay, while $\Phi_{\Gamma^{(k)}}$ represents a MAD where the decays from levels $\ket{k}$ and those above it are forbidden. In this setting, the most generic $d$-dimensional transition matrix $\Gamma$ in \eqref{equation:definition:transitionmatrixold} is equal to $\Gamma ^{(d)}$. By direct computation, setting $k=d-1$ in \eqref{equation:gammadeco1} leads to:
\begin{equation}
	\label{equation:madcompositioniter}
	\Gamma ^{(d)}=\Gamma ^{(d-1)}\Gamma_{d-1},
\end{equation}
then, by iterating \eqref{equation:madcompositioniter}, one arrives at the decomposition:
\begin{equation}
	\label{equation:GammaSingleLevelDeco}
	\Gamma = \Gamma ^{(d)} = \Gamma_1\Gamma_2 \cdots \Gamma_{d-1},
\end{equation}
which, by employing \eqref{equation:MADcomposition}, translates to:
\begin{equation}
	\label{equation:MADdecomposition1}
	\Phi _\Gamma = \Phi_{\Gamma_{d-1}}\circ \cdots \circ \Phi_{\Gamma _1}.
\end{equation}
Intuitively speaking, \eqref{equation:MADdecomposition1} means that in a MAD channel, the lower energy levels "have precedence" when decaying, as represented in \figref{fig:madgraph4}.
\subsubsection{MAD channels as composition of single-decay channels}\label{subsubsection:single-decay}
Consider next the special case of single-decay MAD channels $\Phi_{\Xi _k ^{(n)}}$ characterized by
transition matrices of the form
\begin{equation}
	\label{equation:definition:gammakdecotext}
	\Xi _k ^{(n)} \coloneqq  \mathds{1}+{\xi}_{kn} \op{k}{n} -{\xi}_{kn} \op{k}{k},
\end{equation}
which for $n<k \leq d$, represents single decays from level $\ket{k}$ to $\ket{n}$ with probability
$\xi_{kn}$. It worth observing that the action of $\Phi_{\Xi _k ^{(n)}}$ in the restricted
subspace spanned by the vectors $\ket{k}$ and $\ket{n}$ corresponds to a ADC with loss parameter $\xi_{kn}$.
As shown  in \appref{appendix:isolatedecay}, from~\eqref{equation:GammaSingleLevelDeco}  one can prove that given an arbitrary transition matrix $\Gamma$, setting
\begin{equation}
	\xi_{kn}\coloneqq \label{equation:definition:modifiedamplitudesmaintext}\\
	\frac{\gamma_{kn}}{\gamma_{kn}+(\gamma_{kk}+\sum_{i=0}^{n-1}\gamma_{ki})} \;,
\end{equation}
the following identity holds
\begin{equation}
	\label{equation:GammaADCdeco}
	\Gamma = \prod _{\substack{k = 1\\ \rightarrow}} ^{d-1}\left(\Xi _k ^{(k-1)} \cdots \Xi _k ^{(0)}\right),
\end{equation}
where $\prod _{\rightarrow}$ indicates that the product is meant to be expanded from left to right for increasing $k$'s.
Accordingly we can write
\begin{equation}
	\label{equation:MADADCdeco}
	\Phi _\Gamma = \bigodot _{\substack{k=1\\ \leftarrow}} ^{d-1}\left(\Phi_{\Xi _k ^{(0)}}\circ \cdots \circ \Phi_{\Xi _k ^{(k-1)}}\right),
\end{equation}
where now $\bigodot _{\leftarrow}$ indicates a composition of channels that is meant to be expanded from right to left for increasing $k$'s.
\subsection{Conjugation with unitary transformations}
\label{unitarycovariancecapacity}

Many MAD's which are seemingly different but whose decay probabilities have the same numerical values actually possess the same communication capabilities up to unitary operations acting on the input and the output of the process.
Indeed given two transition matrices $\Gamma'$ and $\Gamma''$ unitarily connected via
a permutation matrix $P$, it follows that the  corresponding MAD channels are unitary equivalent:
\begin{equation} \label{equation:unitaryconjugation:gamma}
	\Gamma'' = P \Gamma' P^{\dag} \quad \Rightarrow \quad
	\Phi_{ \Gamma''}=\mathcal{P}^{\dag}\circ\Phi_{ \Gamma'}\circ \mathcal{P}\;,
\end{equation}
where   $\mathcal{P}(\cdots):=P (\cdots )P^{\dagger}$.
This for instance applies to all single-decay MAD channels $\Phi_{ \Xi_{k_1}^{(n_1)} }$ and $\Phi_{\Xi _{k_2} ^{(n_2)}}$
which share the same value  ($\bar{\Omega}_{k_1 n_1} = \bar{\Omega}_{k_2 n_2}$) of the transition parameters.

We further recall that
MAD channels are covariant under conjugation with unitary transformations $U$ that are diagonal w.r.t. the computational basis~\cite{mad3} , i.e.
\begin{eqnarray} \nonumber
	U|j\rangle &=e^{i\varphi_j}| j\rangle \quad \forall j \in \{1,\cdots, d\}\\ &\Longrightarrow\label{unitaryconjugation} \;
	\Phi_{\Gamma}=\mathcal{U}\circ\Phi_{\Gamma}\circ \mathcal{U}^{\dagger}\;,
\end{eqnarray}
with $\mathcal{U}(\cdots)=U (\cdots )U^{\dagger}$.

\subsection{Complementary channels}
Complementary channels express how information is lost to the external environment during a noisy quantum evolution~\cite{StinespringDilation}.
In the case of a MAD channel $\Phi_{\Gamma}$, the explicit expression of the associated complementary channel
$\tilde{\Phi}_{\Gamma}$ can be obtained in terms of the Kraus
representation~(\ref{equation:definition:madchannelkraus}).
Specifically, following
~\cite{holevo_book}, we get
\begin{eqnarray}
	\label{equation:ComplMAD0}
	\tilde{\Phi}_{\Gamma}(\rho)&=& \tr\left[K_{00}\rho K^\dagger_{00}\right] \op{0,0}{0,0}\\ \nonumber
	&&+\sum_{i<j}\left( \tr\left[K_{00}\rho K^\dagger_{ij}\right] \op{0,0}{i,j}+\mathrm{h.c.}\right)\\
	\nonumber
	&&+ \sum_{i<j}\sum_{i'<j'} \tr\left[K_{ij}\rho K^\dagger_{i'j'}\right] \op{i,j}{i',j'}\\
	\label{equation:ComplMAD}
	&=&\sum_{j=0}^{d-1}\gamma_{jj}\rho_{jj} \op{0,0}{0,0}\\ \nonumber
	&&+\sum_{i<j}\left(\sqrt{\gamma_{ii}\gamma_{ji}}\rho_{ij} \op{0,0}{i,j}+\mathrm{h.c.}\right)\\ \nonumber
	&&+ \sum_{i<j,j'}\sqrt{\gamma_{ji}\gamma_{j'i}}\rho_{jj'}\op{i,j}{i,j'},
\end{eqnarray}
where we set  $\sum_{i<j} =  \sum_{j=1}^{d-1} \sum_{i=0}^{j-1}$ for ease of notation, and where the vectors
$\left\{\ket{0,0},\left\{\ket{i,j}\right\}_{0\leq i<j\leq d-1}\right\}$ represent an orthonormal basis for the system environment.

\subsection{Inverse maps of MAD channels}

The inverse of a channel $\Phi$ is defined as the (not necessarily LCPTP) super-operator $\Phi^{-1}$ such that
\begin{eqnarray}
	\Phi \circ \Phi^{-1} = \Phi^{-1} \circ \Phi = \mbox{Id} \; .\label{definition:inverse}
\end{eqnarray}
Not all LCPTP maps admit an inverse: but if a channel $\Phi$ does, we say that is mathematically invertible, or simply invertible.
In the case of  MAD channels one can explicitly derive  necessary and sufficient conditions for invertibility
in terms of the matrix elements of the transition matrices, and provide an explicit expression for the associated inverse.
\\

Let first consider the  case of a single-decay MAD channels $\Phi_{\Xi _k ^{(n)}}$ introduced in
Sec.~\ref{subsubsection:single-decay}. It turns out that as long as the probability parameter $\xi_{kn}$ which defines $\Xi _k ^{(n)}$ as in Eq.~(\ref{equation:definition:gammakdecotext})  is strictly smaller than $1$ (i.e. $\xi_{kk}:= 1-\xi_{kn} > 0$) , then $\Phi_{\Xi _k ^{(n)}}$ is invertible with inverse given by the super-operator
\begin{equation}
	\label{equation:inverseMADsingledecay}
	\Phi ^{-1} _{\Xi _k ^{(n)}} (\cdots) \coloneqq \tilde{K}_{k;0} ^{(n)} (\cdots) \tilde{K}_{k;0} ^{(n)^\dagger} -\tilde{K}_{k;1} ^{(n)} (\cdots) \tilde{K}_{k;1} ^{(n)^\dagger} \;, \end{equation}
with
\begin{eqnarray}
	\tilde{K}_{k;0} ^{(n)}  &\coloneqq& \mathds{1} -\left(1- \frac{1}{\sqrt{1-\xi_{kn}}}\right) \op{k}{k},\nonumber\\
	\tilde{K}_{k;1} ^{(n)}  &\coloneqq& \sqrt{\frac{\xi_{kn}}{1-\xi_{kn}}}\op{n}{k},
\end{eqnarray}
The importance of \eqref{equation:inverseMADsingledecay} lies in the fact that it can be composed to generate the inverse map of a general MAD channel. In fact, recall that in \eqref{equation:MADADCdeco} it was shown that any MAD channel
$\Phi _{\Gamma}$  can be seen as a composition of single-decay MAD channels. Then, by "inverting" those single-decays one by one, the resulting channel must be the identity channel. This line of reasoning results in the definition:
\begin{equation}
	\label{equation:definition:MADinverse}
	\Phi _{\Gamma} ^{-1} = \bigodot _{\substack{k=1\\ \rightarrow}} ^{d-1}\left(\Phi ^{-1} _{\Xi _k ^{(k-1)}} \circ \cdots \circ \Phi ^{-1} _{\Xi _k ^{(0)}} \right)\;,
\end{equation}
which is well defined as long as all the single-decay rates of the maps $\Phi _{\Xi _k ^{(n)}}$ which enter the decomposition are strictly smaller than 1. Recalling Eq.~(\ref{equation:definition:modifiedamplitudesmaintext}) this translates in the  constraint
\begin{eqnarray} \label{invertibilitycondition}
	\gamma_{kk} > 0\;, \quad \forall k\in \{0, \cdots, d-1\}\;,
\end{eqnarray}
which is the necessary and sufficient condition for the invertibility of $\Phi_{\Gamma}$.
Furthermore since,  ${\Gamma}$ is by construction lower-triangular Eq.~\eqref{invertibilitycondition}  translates into the strict positivity of its determinant, hence into its invertibility, leading us to the condition
\begin{equation} \label{invertibilitycondition11}
	\mbox{$\Phi_\Gamma$ invertible}  \quad \Longleftrightarrow \quad  \mbox{$\Gamma$ invertible}\;.
\end{equation}
Observe finally that   since the inverse of a lower-triangular matrix is also lower-triangular, we can write
\begin{equation} \label{invertibilitycondition111}
	\Phi^{-1}_\Gamma=\Phi_{\Gamma^{-1}}\;,
\end{equation}
where $\Phi_{\Gamma^{-1}}$ is the (not-necessarily LCPTP) map defined as in \eqref{equation:definition:madchannelkraus1} replacing the matrix
elements of $\Gamma$ with the corresponding ones of $\Gamma^{-1}$.
We recall that a Choi state of quantum channel $\Phi$ is obtained applying
the extension of the map ${\text Id} \otimes \Phi $  on a maximally entangled input state $|\Psi_{\max}\rangle$ of~${\cal H}\otimes {\cal H}$.
Identifying   $|\Psi_{\max}\rangle$ with the maximally correlated state w.r.t. to the computational basis
($\sum_{j=0}^{d-1} |j\rangle_A \otimes |j\rangle_B/{\sqrt d}$) the Choi state of the MAD channel $\Phi_{\Gamma}$   can be expressed as
\begin{eqnarray}
	\label{equation:madchoi}
	\rho_{AB}^{(\Phi_{\Gamma})}&=&  \frac{1}{d} \sum_{j=0}^{d-1}
	\sum_{i=0}^{j}\gamma_{ji} |j\rangle_{A}\langle j |\otimes |i\rangle_B\langle i|
	\\ &&+\frac{1}{d} \sum_{\substack{i,j=0 \\ i\neq j}}^{d-1}\sqrt{\gamma _{jj}\gamma_{ii}}|j\rangle_{A}\langle i|\otimes |j\rangle_B\langle i|, \nonumber
\end{eqnarray}
where we used the indexes $A$ and $B$ to label the two copies of ${\cal H}$.

\section{Degradability analysis}
A quantum channel $\Phi$ is said to be \textit{degradable} if there exists a LCPTP map $\Lambda$ such that the composition $\Lambda\circ \Phi$ is equal to the complementary channel $\tilde{\Phi}$ of $\Phi$, i.e. $\tilde{\Phi} = \Lambda\circ {\Phi}$.
Conversely, $\Phi$ is said to be \textit{antidegradable} if there exists a LCPTP map $\Lambda ^\prime$ such that the composition $\Lambda^\prime\circ \tilde{\Phi}$ is equal to the channel $\Phi$, i.e.
$\Phi = \Lambda^\prime\circ\tilde{\Phi}$. It turns out that when either one of these properties holds, the formula which expresses the optimal rate for reliably transmitting quantum
information
using the channel $\Phi$, i.e. its
quantum capacity $Q(\Phi)$, simplifies~\cite{qcapacitydef}.
Indeed, on one hand, the quantum capacity of antidegradable channels,  is exactly zero:
\begin{equation}
	\label{equation:qcapacityantideg}
	\mbox{$\Phi$ antideg.}\;  \Rightarrow \; 	Q\left(\Phi\right) = 0
	\;.  		\end{equation}
We also remark that, given $\Phi$  antidegradable and $\Phi'$ an arbitrary LCPTP map, the quantum capacities  of
$\Phi\circ \Phi'$ and $\Phi'\circ \Phi$ are both zero (in particular $\Phi\circ \Phi'$ is also antidegradable), i.e.
\begin{equation}
	\label{equation:qcapacityantideg1}
	\mbox{$\Phi$ antideg.}\;  \Rightarrow \; 	Q(\Phi\circ \Phi') = Q(\Phi'\circ \Phi) =0
	\;.  		\end{equation}
On the other hand,
the coherent information of degrable channels
\begin{equation} \label{coherentinfo}
	I_c (\rho ,\Phi):= S(\Phi(\rho)) - S(\tilde{\Phi}(\rho))\;,\end{equation} is additive and concave
with respect to the input state $\rho$ (here $S(\cdots) := -\tr[ (\cdots) \log_2 (\cdots)]$ is the von Neumann entropy). This implies that $Q\left(\Phi\right)$ can be expressed as a single-letter formula obtained by  maximizing
$I_c (\rho ,\Phi)$ over $\rho$.
For degradable MAD channels this further simplifies thanks to the covariance property~(\ref{unitaryconjugation}) which allows one to restrict the maximization on  the set of density matrices which are
diagonal w.r.t. the computational basis~\cite{mad3}, leading to
\begin{equation}
	\label{equation:qcapacitydegMAD}
	\mbox{$\Phi_\Gamma$ deg.}	\;  \Rightarrow \; Q\left(\Phi_\Gamma\right) = \max _{\substack{\rho \in\mathfrak{S}\left(\mathcal{H}\right) \\\rho \textrm{ diagonal}}}I_c \left(\rho ,\Phi_\Gamma\right).
\end{equation}
Notice also that if the composite channel $\Phi \circ \Phi'$ is degradable, then $\Phi$ is also degradable. Furthermore if $\Phi$ is invertible (see \eqref{definition:inverse}), $\Phi'$ is degradable as well:
\begin{eqnarray}
	\label{equation:qcapacitydeg1}
	&\!\!\!\!\mbox{$\Phi\circ \Phi'$ deg.}\;  \Rightarrow \; 	\mbox{$\Phi'$ deg.}& \\
	&\!\!\!\!\mbox{$\Phi\circ \Phi'$ deg. $\&$ $\Phi'$ invertible}  \;  \Rightarrow \; 	\mbox{$\Phi$ deg.}&
	\label{equation:qcapacitydeg2}	\end{eqnarray}
While Eqs.~(\ref{equation:qcapacityantideg1}) and (\ref{equation:qcapacitydeg1}) are well known relations (see e.g. ~\cite{holevo_book}), we are not aware of a derivation of (\ref{equation:qcapacitydeg2}) so we report it in Appendix~\ref{appendix:degradabledecomposition}. \\

\subsection{Anti-degradable MAD channels}
In this section we present a complete characterization of the antidegradability region in the parameter space of $d$-dimensional MAD channels.
Coincidentally, we show that any MAD channel that does not meet these conditions has quantum capacity strictly larger than $0$, which means that those that do meet them are the only "useless" ones in terms of communication capabilities.
\\
\begin{theorem}\label{theorem1}
	A MAD channel $\Phi _{\Gamma}$ is anti-degrabable iff the
	elements  of its transition matrix $\Gamma$
	fulfil the inequalities
	\begin{equation}
		\label{equation:antidegconditions}
		\gamma_{j0}-\gamma_{jj}\geq 0 \qquad \forall j=1,\cdots,d-1.
	\end{equation}
	Furthermore the quantum capacity of MAD channel $\Phi_{\Gamma}$  is zero if and only if
	it is anti-degradable,
	\begin{equation}
		\label{equation:zerocapacitycondition}
		Q\left(\Phi _{\Gamma}\right) = 0\quad \Longleftrightarrow \quad \mbox{$\Phi_{\Gamma}$ {\rm antideg.}}
	\end{equation}
\end{theorem}
We identified the conditions given in \eqref{equation:antidegconditions}  from results obtained for $3,4$-dimensional MAD's, supported by
numerical exploration  of the two-extendibility criterion for the Choi state of the channel~\cite{2extension}. This was done through semi-definite programming \cite{SDP} (implemented in \texttt{Python}).
In the case at hand, we want to find out a
whether there exists a three-partite state
that satisfies the conditions in \eqref{ext.cond} for a certain set of parameters $\gamma_{ji}$. In \listref{lst:semidef} the basic structure of the function used for implementing the semidefinite program is reported. Inferring from this numerical process, an analytical candidate for the two-extension of the Choi states of antidegradable MAD's was obtained. 			\begin{lstlisting}[language=Python, caption=Function to check if a Choi state \texttt{choi} is $2$-extendable, label={lst:semidef}]
		def Extendability(choi,d):
			X = cvxpy.Variable((d**3, d**3), hermitian=True) # Extension
			constraints = [] # List of constraints
			objective = cvxpy.Minimize(np.sum(X.value)) # Might set a lot of matrix elements to 0
			constraints.append(X>>0)
			constraints.append(np_array_as_expr(partial_trace(expr_as_np_array(X),sys=[2],dim=3*[d]))==choi)
			constraints.append(np_array_as_expr(partial_trace(expr_as_np_array(X),sys=[1],dim=3*[d]))==choi)
			
			print("START!\n")
			
			prob = cvxpy.Problem(objective,constraints) # Initialize problem
			prob.solve(solver=cvxpy.MOSEK,verbose=False) # Call solver
			
			if prob.status not in ["infeasible","unbounded"]:
			print('Antidegradable.\n')
			print(np.real(np.round(X.value,2)))
			return 1
			else:
			return 0
		\end{lstlisting}
\begin{proof}
	As a first step we observe that, thanks to~(\ref{equation:qcapacityantideg}), proving Theorem~\ref{theorem1} reduces to demonstrate two statements, i.e.
	\begin{eqnarray}
		\label{equation:antidegsuff}
		\!\! \!\!\{ \gamma_{j0}-\gamma_{jj}\geq 0 \quad \forall j \} &\Longrightarrow&  \mbox{$\Phi_{\Gamma}$ {\rm antideg.}}\;, \\
		\{\exists j: \gamma_{j0}<\gamma_{jj}\} &\Longrightarrow& Q\left(\Phi_\Gamma\right) > 0\;.\label{equation:antidegnec}
	\end{eqnarray}
	\paragraph*{Part I:--}
	Here we prove Eq.~(\ref{equation:antidegsuff}), i.e. that  (\ref{equation:antidegconditions}) is a sufficient condition  for
	anti-degradability of MAD channels.
	We recall that a necessary and sufficient condition for anti-degradability is that the corresponding Choi state of the channel
	is \emph{two-extendible}~\cite{2extension}, i.e.
	\begin{equation}
		\mbox{$\Phi_\Gamma$ antideg.}\;  \Leftrightarrow
		\label{equation:2extantideg}
		\mbox{$\rho_{AB}^{(\Phi_{\Gamma})}$ two-extendible.}
	\end{equation}
	Recall also that the two-extensibility property of $\rho_{AB}^{(\Phi_{\Gamma})}$ means that
	there exists a density matrix $\rho^{(\rm{ext})}_{AB_1B_2}$ on three copies of ${\cal H}$ (i.e.
	${\cal H}\otimes {\cal H}\otimes {\cal H
		}$) whose reduced density matrices on $AB_1$ and $AB_2$  provides 	copies of 	$\rho_{AB}^{(\Phi_{\Gamma})}$, i.e.
	\begin{eqnarray}
		\left\{ \begin{array}{l}
			\rho^{(\rm{ext})}_{AB_1}:=\tr _{B_2}[ \rho^{(\rm{ext})}_{AB_1B_2}] = \rho^{(\Phi_{\Gamma})}_{AB_1}\;, \\
			\label{ext.cond}                                                                                      \\
			\rho^{(\rm{ext})}_{AB_2}:=\tr _{B_1} [\rho^{(\rm{ext})}_{AB_1B_2}]= \rho^{(\Phi_{\Gamma})}_{AB_2}\;,\end{array} \right.
	\end{eqnarray}
	Starting from the expression~(\ref{equation:madchoi}) of $\rho_{AB}^{(\Phi_{\Gamma})}$, numerical
	analysis suggested  that a suitable candidate for $\rho^{(\rm{ext})}_{AB_1B_2}$ is provided by
	operators of the form
	\begin{equation}
		\label{extension}
		\rho^{(\rm{ext})}_{AB_1B_2} = \tau^{(\rm diag)} _{AB_1B_2} +\tau^{(\rm off)} _{AB_1B_2} .
	\end{equation}
	Here $\tau^{(\rm diag)} _{AB_1B_2}$ is an operator which is diagonal w.r.t. to $A$ and has the form
	\begin{equation}
		\tau^{(\rm diag)} _{AB_1B_2} \coloneqq\frac{1}{d}\sum_{j=0} ^{d-1}|j\rangle_{A}\langle j | \otimes \tau_{B_1B_2} ^{(j)},
	\end{equation}
	with $\tau_{B_1B_2} ^{(j)}$   symmetric under exchange of $B_1$ and $B_2$
	(i.e.  $\tau_{B_1B_2} ^{(j)}=\tau_{B_2B_1} ^{(j)}$), defined as
	\begin{eqnarray}
		&&\tau_{B_1B_2}^{(j)}\coloneqq
		\sum_{i=1}^{j-1}\gamma_{ji}|i\rangle_{B_1}\langle i|\otimes|i\rangle_{B_2}\langle i|\\
		&&+ \gamma_{jj}\left(
		|0\rangle_{B_1}\langle 0|\otimes |j\rangle_{B_2}\langle j |+|j\rangle_{B_1}\langle j| \otimes|{0}\rangle_{B_2}\langle 0|\right)\nonumber \\
		&&+\gamma_{jj}\left(1-\delta_{j0}\right)\left(|{0}\rangle_{B_1}\langle j |\otimes |j\rangle_{B_2}\langle 0|+\mbox{h.c.}\right)\nonumber \\
		&&+\left(\gamma_{j0}-\gamma_{jj}\right) |0\rangle_{B_1}\langle 0| \otimes \Big(\sum_{i=0}^{j-1}p_i^{(j)}|i\rangle_{B_2}\langle i|\Big)\nonumber \\
		&&+(\gamma_{j0}-\gamma_{jj})\Big( \sum_{i=1} ^{j-1}
		p^{(j)}_i	|i\rangle_{B_1}\langle i|\Big) \otimes| 0\rangle_{B_2}\langle 0| \nonumber \\
		&&-(\gamma_{j0}-\gamma_{jj})\Big( \sum_{i=1} ^{j-1}
		p^{(j)}_i	|i\rangle_{B_1}\langle i| \otimes |i\rangle_{B_2}\langle i|\Big), \nonumber
	\end{eqnarray}
	where $p_i^{(j)}$  are arbitrary positive constants which for fixed $j$ define a probability distribution w.r.t. to the $i$ index, i.e.
	\begin{eqnarray} \left\{ \begin{array}{ll}
			p^{(j)}_i \in [0,1]\;, \qquad \forall i\in \{ 0,1,\cdots,j-1\}\;, \\\\
			\sum_{i=0}^{j-1} p^{(j)}_i =1\;. \label{constraintofp}
		\end{array} \right.
	\end{eqnarray}

	The other contribution in \eqref{extension} is instead off-diagonal in $A$:
	\begin{equation}
		\tau^{(\rm off)} _{AB_1B_2} \coloneqq \frac{1}{d} \sum _{\substack{i,j=0 \\ i\neq j}}
		\sqrt{\gamma_{jj}\gamma_{ii}} |j\rangle_{A}\langle i| \otimes \tau_{B_1B_2} ^{(ji)},
	\end{equation}
	with $\tau_{B_1B_2} ^{(ji)}$ operators symmetric under exchange of $B_1$ and $B_2$  defined as
	\begin{eqnarray}\nonumber
		\tau_{B_1B_2} ^{(ji)} &\coloneqq&	|0\rangle_{B_1}\langle 0|\otimes | j\rangle_{B_2}\langle i|+	| j\rangle_{B_1}\langle i|\otimes | 0\rangle_{B_2}\langle 0|\\
		&&+\left(1-\delta_{j0}\delta_{i0} \right)\Big(| j\rangle_{B_1}\langle 0|\otimes | 0\rangle_{B_2}\langle i|\nonumber \\
		&&\qquad
		\qquad + | 0\rangle_{B_1}\langle i|\otimes | j\rangle_{B_2}\langle 0|\Big).
	\end{eqnarray}
	Observe that, for all choices of the coefficients $p_i^{(j)}$ that fulfil the normalisation constraint~\eqref{constraintofp}, the partial traces over the subspaces $B_1,B_2$ of $\rho^{(\rm{ext})}_{AB_1B_2}$ return the desired $\rho^{(\Phi_{\Gamma})}_{AB_1}$ and $\rho^{(\Phi_{\Gamma})}_{AB_2}$.  Therefore  the only thing left to be proven is that $\rho^{(\rm{ext})}_{AB_1B_2}$ is actually a state, which requires checking its positivity.

	Looking at \eqref{extension}, it is possible to recognize that, through  a unitary permutation of the basis elements,  $\rho^{(\rm{ext})}_{AB_1B_2}$ can be cast in the diagonal block form
	\begin{equation}\label{matrixformextchoi}
		\rho^{(\rm{ext})}_{AB_1B_2}= \left[\begin{array}{cc}
				M & 0 \\
				0 & D
			\end{array}\right],
	\end{equation}
	which allows us to reduce the positivity  analysis of $\rho^{(\rm{ext})}_{AB_1B_2}$
	into  the positivity analysis of $M$ and $D$.
	The block $D$ is a diagonal matrix that includes
	contributions of $\rho^{(\rm{ext})}_{AB_1B_2}$ which can be directly
	associated to
	explicit eigenvectors of the operator: i.e.
	\begin{equation}
		\label{diagonal0}
		\left\{
		\begin{array}{l}
			\gamma_{ji}| j\rangle_{A}\langle j |\otimes |i\rangle_{B_1}\langle i|\otimes
			|i\rangle_{B_2}\langle i|,
			\\\\
			p^{(j)}_i(\gamma_{j0}-\gamma_{jj})| j\rangle_{A}\langle j |\otimes|0\rangle_{B_1}\langle 0|\otimes|i\rangle_{B_2}\langle i|,
			\\\\
			p^{(j)}_i(\gamma_{j0}-\gamma_{jj}) | j\rangle_{A}\langle j |\otimes|i\rangle_{B_1}\langle i|\otimes| 0\rangle_{B_2}\langle 0|,
			\\\\
			(\gamma_{ji}-p^{(j)}_i(\gamma_{j0}-\gamma_{jj})) \\\\
			\qquad \qquad \times | j\rangle_{A}\langle j |\otimes|i\rangle_{B_1}\langle i|\otimes|i\rangle_{B_2}\langle i|,\end{array} \right.
	\end{equation}
	$\forall j=1,\cdots,d-1$ and $\forall i=0,\cdots,j-1$.
	The  block $M$ includes the remaining terms of
	$\rho^{(\rm{ext})}_{AB_1B_2}$ and can be casted in the matrix form
		{\footnotesize
			\begin{equation}
				\left(\begin{array}{cccccc}
						1                  & \sqrt{\gamma_{11}}            & \sqrt{\gamma_{11}}            & \sqrt{\gamma_{22}}            & \sqrt{\gamma_{22}}            & \cdots \\
						\sqrt{\gamma_{11}} & \gamma_{11}                   & \gamma_{11}                   & \sqrt{\gamma_{11}\gamma_{22}} & \sqrt{\gamma_{11}\gamma_{22}} & \cdots \\
						\sqrt{\gamma_{11}} & \gamma_{11}                   & \gamma_{11}                   & \sqrt{\gamma_{11}\gamma_{22}} & \sqrt{\gamma_{11}\gamma_{22}} & \cdots \\
						\sqrt{\gamma_{22}} & \sqrt{\gamma_{11}\gamma_{22}} & \sqrt{\gamma_{11}\gamma_{22}} & \gamma_{22}                   & \gamma_{22}                   & \cdots \\
						\sqrt{\gamma_{22}} & \sqrt{\gamma_{11}\gamma_{22}} & \sqrt{\gamma_{11}\gamma_{22}} & \gamma_{22}                   & \gamma_{22}                   & \cdots \\
						\vdots             & \vdots                        & \vdots                        & \vdots                        & \vdots                        & \ddots
					\end{array}\right). \label{A}
			\end{equation}}
	Proving  the positivity of the  block $M$ requires making use of Sylvester's criterion:
	$M\geq 0$ if and only if all its \text{principal minors}  are non-negative. Looking at \eqref{A}, it is possible to notice that many of the principal minors will be $0$; in fact, every time the submatrix obtained from $M$ by "cutting" some rows (and corresponding columns) contains two copies of the same row, its determinant is automatically $0$. This means that in order to prove the positivity of $M$, we only need to prove the positivity of just one of its submatrices, specifically:
	\begin{equation}
		\label{Atilde}
		\tilde{M}=\left(\begin{array}{cccc}
				1                  & \sqrt{\gamma_{11}}            & \sqrt{\gamma_{22}}            & \cdots \\
				\sqrt{\gamma_{11}} & \gamma_{11}                   & \sqrt{\gamma_{11}\gamma_{22}} & \cdots \\
				\sqrt{\gamma_{22}} & \sqrt{\gamma_{11}\gamma_{22}} & \gamma_{22}                   & \cdots \\
				\vdots             & \vdots                        & \vdots                        & \ddots
			\end{array}\right).
	\end{equation}
	Computing the expectation value of $\tilde{M}$ on a generic pure state $\ket{\psi}= \sum _i\alpha_i|i\rangle$ in the relevant $d$-dimensional subspace of the Hilbert space $\mathcal{H}_A\otimes\mathcal{H}_{B_1}\otimes\mathcal{H}_{B_2}$ leads to:
	\begin{eqnarray} \nonumber
		\ev{\tilde{M}}{\psi} &=&(\sum_{i=0}^{d-1}\sqrt{\gamma_{ii}})\langle \psi| \sum _i\gamma_{ii} \alpha_i|i\rangle \\
		&=&\left|\sum_{i=0}^{d-1}\sqrt{\gamma_{ii}}\alpha_i\right|^2\geq 0 \;,
	\end{eqnarray}
	for all $|\psi\rangle$.

	Since $D$ is already  in diagonal form, its positivity condition results  in imposing  all its diagonal elements  to be non-negative:
	\begin{eqnarray}
		\gamma_{ji} &\geq& 0\;, \label{diagonal1} \\
		p^{(j)}_i(\gamma_{j0}-\gamma_{jj}) &\geq& 0\;,   \label{diagonal2}\\
		\gamma_{ji}-p^{(j)}_i(\gamma_{j0}-\gamma_{jj})&\geq& 0\;, \label{diagonal3}
	\end{eqnarray}
	$\forall j=1,\cdots,d-1$ and $\forall i=0,\cdots,j-1$.
	The first condition is ensured by construction.
	Recalling that  the coefficients $p^{(j)}_i$ are positive semidefinite (see \eqref{constraintofp}),
	Eq.~(\ref{diagonal2}), instead, imposes to have
	\begin{equation}
		\label{antidegcond}
		\gamma_{j0}-\gamma_{jj}\geq 0 \quad \quad \forall j=1,\cdots,d-1.
	\end{equation}
	Finally, under  \eqref{antidegcond},  the condition \eqref{diagonal3}  requires us to show that among the allowed coefficients $p^{(j)}_i$ we can find some such that
	\begin{equation} \label{diagonal3new}
		p_{i}^{(j)} \leq \frac{\gamma_{ji}}{\gamma_{j0}-\gamma_{jj}}\qquad \forall  i <j\;.
	\end{equation}
	Recalling that for fixed $j$, the coefficient $\gamma_{ji}$ define   a probability distribution for $i=0,1,\cdots,j$, such solution can be found e.g. by taking
	\begin{equation} \label{diagonal3newsol}
		p_{i}^{(j)}= \frac{\gamma_{ji}}{1-\gamma_{jj}}\qquad \forall  i <j\;.
	\end{equation}
	This allows us to conclude that the block $D$, and hence of the entire operator
	$\rho^{(\rm{ext})}_{AB_1B_2}$ is positive semidefinite whenerver
	\eqref{antidegcond} is fulfilled.\\
	\paragraph*{Part II:--} Here we prove Eq.~(\ref{equation:antidegnec}), i.e. that
	(\ref{equation:antidegconditions}) is a necessary condition  to have a non-zero quantum capacity value.

	Given the MAD channel $\Phi_{\Gamma}$ and $j\in \{1,\cdots, d-1\}$ consider its restriction
	$\Phi^{[0,j]}_{\Gamma}$ which  acts on
	inputs states associated to the subspace
	$\mathcal{H}^{[0,j]}\coloneqq\mathrm{Span}\left\{| 0\rangle,| j\rangle\right\}$.
	By construction $\Phi^{[0,j]}_{\Gamma}$ has the same output as the original MAD channel
	$\Phi_{\Gamma}$ when the latter is applied on inputs lying in~$\mathcal{H}^{[0,j]}$, which leads us to
	the inequality
	\begin{eqnarray} \label{restriction}
		Q\left(\Phi_{\Gamma}\right) &\geq &Q\left(\Phi^{[0,j]}_{\Gamma}\right) \;.			\end{eqnarray}
	From \eqref{equation:definition:MADkrausset} we can derive a Kraus decomposition for 	$\Phi^{[0,j]}_{\Gamma}$ formed by  $j+1$ operators, represented by $(j+1)\times 2$ matrices:
	\begin{eqnarray} \nonumber
		{K}^{[0,j]}_{i} &\coloneqq& {K}_{ij} =\sqrt{\gamma_{ji}}\op{i}{j} \quad 0\leq i<j\;, \\
		{K}^{[0,j]}_{00}  &\coloneqq& \op{0}{0} +\sqrt{\gamma_{jj}}\op{j}{j}\;.
	\end{eqnarray}
	Let  then $\rho = (1-p)\op{0}{0} +p\op{j}{j} +\lambda
		\op{0}{j} + \lambda^* \op{j}{0}$ be a density matrix of $\mathcal{H}^{[0,j]}$.
	The corresponding output state
	$\Phi^{[0,j]}_{\Gamma}(\rho) = (1-p) \op{0}{0} + \sum _{i =0} ^{j}\gamma_{ji}p\op{i}{i}  +\left(\sqrt{\gamma_{jj}}\lambda\op{0}{j} +\textrm{h.c.}\right)$ generated by the restriction of $\Phi_{\Gamma}$
	can be cast in the matrix form
		{\footnotesize\begin{eqnarray}
				&& \nonumber \begin{pmatrix}
					(1-p) +\gamma_{j0} p         & 0             & 0             & \ldots & 0                & \sqrt{\gamma_{jj}}\lambda \\
					0                            & \gamma_{j1} p & 0             & \dots  & 0                & 0                         \\
					0                            & 0             & \gamma_{j2} p &        & \vdots           & \vdots                    \\
					\vdots                       & \vdots        &               & \ddots &                  &                           \\
					0                            & 0             & \dots         &        & \gamma_{j,j-1} p & 0                         \\
					\sqrt{\gamma_{jj}}\lambda ^* & 0             & \dots         &        & 0                & \gamma_{jj} p             \\
				\end{pmatrix}.
			\end{eqnarray}}
	Using the notation introduced in \eqref{equation:ComplMAD}, the output of the complementary channel $\Phi^{[0,j]}_{\Gamma}$ on $\rho$ is instead the density matrix
	\begin{eqnarray}
		&&\tilde{\Phi}^{[0,j]}_{\Gamma}(\rho)  =( (1-p)+\gamma_{jj} p) \op{0,0}{0,0} \\
		&& \;+\sum _{i =0} ^{j-1}\gamma_{ji}p\op{i,j}{i,j} +\left(\sqrt{\gamma_{j0}}\lambda\op{0,0}{0,j} +\textrm{h.c.}\right) \nonumber
	\end{eqnarray}
	which can be written as the following $j\times j$ block matrix
		{\footnotesize \begin{equation} \nonumber
				\left(\begin{array}{cc|ccc}
					(1-p) +\gamma_{jj} p          & \sqrt{\gamma_{j0}} \lambda &               &        &                  \\
					\sqrt{\gamma_{j0}} \lambda ^* & \gamma_{j0}p               &               &        &                  \\ \hline
					                              &                            & \gamma_{j1} p &        &                  \\
					                              &                            &               & \ddots &                  \\
					                              &                            &               &        & \gamma_{j,j-1} p \\
				\end{array}\right).
			\end{equation}}
	Observe next that
	under the condition
	\begin{equation}\gamma_{j0}<\gamma_{jj},\label{condition}\end{equation}
	the channle  $\Phi^{[0,j]}_{\Gamma}$ is degradable with  degrading map given by  the single-decay MAD channel $\Phi_{\Xi _j ^{(0)}}$
	with decay probability
	${\xi}_{j0} \coloneqq \frac{\gamma_{jj}-\gamma_{j0}}{\gamma_{jj}}$, i.e.
	\begin{equation}
		\Phi_{\Xi _j ^{(0)}} \circ\Phi^{[0,j]}_{\Gamma} = \tilde{\Phi}^{[0,j]}_{\Gamma}\,.
	\end{equation}
	Observing finally that $\Phi^{[0,j]}_{\Gamma}$ inherites from
	$\Phi_{\Gamma}$ the  covariant property~\eqref{unitaryconjugation}, we can invoke \eqref{equation:qcapacitydegMAD} to express the quantum capacity of $\Phi^{[0,j]}_{\Gamma}$ as a maximization of its coherent
	information $ I_c (\rho ,\Phi^{[0,j]}_{\Gamma})$ over the set of diagonal input states, i.e. 			 		\begin{eqnarray}
		&&\!\!\!\!\!\!\!\!\!\!Q\left(\Phi^{[0,j]}_{\Gamma}\right) = \max _{\substack{\rho \in\mathfrak{S}\left(\mathcal{H}^{[0,j]}\right) \\\rho \textrm{ diagonal}}}I_c \left(\rho ,\Phi^{[0,j]}_{\Gamma}\right) \\ \nonumber
		&&=\max_{p\in [0,1]} \Big\{- \gamma_{jj} p\log_2\left(\gamma_{jj}p\right)+\gamma_{j0} p\log_2(\gamma_{j0}p)
		\\\nonumber
		&&-\left((1-p)+\gamma_{j0} p\right)\log _2 \left((1-p)+\gamma_{j0} p\right)\\
		&&+\left((1-p)+\gamma_{jj} p\right)\log _2 \left((1-p)+\gamma_{jj} p\right)\Big\} \;.\nonumber
	\end{eqnarray}
	The last expression can be numerically solved: for our purposes however we need only to verify that $Q\left(\Phi^{[0,j]}_{\Gamma}\right)$ is strictly positive when \eqref{condition} holds.  For this  we set $p=1/2$ obtaining the lower bound
	\begin{eqnarray} \label{lowerboundforQ}
		Q\left(\Phi^{[0,j]}_{\Gamma}\right) &\geq &\log_2\sqrt{\frac{f(\gamma_{j0})}{f(\gamma_{jj})}}\;,			\end{eqnarray}
	with \begin{equation}
		f(x) \coloneqq \frac{x^x}{(1+x)^{(1+x)}},\qquad 0\leq x\leq 1 \;.
		\label{func}
	\end{equation}
	Since the first derivative of such a function $f'(x) = f(x)\log_2 \left(\frac{x}{1+x}\right)$  is strictly positive in the entire domain, we can conclude that
	under the condition \eqref{condition} the right-hand-side term of
	Eq.~\eqref{lowerboundforQ}
	is strictly positive.  Accordingly we can conclude that
	\begin{equation}
		\gamma_{j0}<\gamma_{jj} \Longrightarrow Q\left(\Phi^{[0,j]}_{\Gamma}\right) > 0
		\Longrightarrow
		Q\left(\Phi_{\Gamma}\right) > 0\;, \label{equation:antidegnec1}
	\end{equation}
	which concludes the proof.
\end{proof}
\subsection{Degradability analysis of MAD channels}
\label{section:degradability}
\begin{figure}
	\centering
	\begin{tikzpicture}[scale=1.5]
	\draw[black, thick] (5.5,0) -- (3.5,0) node [left] {\small{$\ket{0}$}};
	\draw[black, thick] (5.5,0.5) -- (3.5,0.5) node [left] {\small{$\ket{1}$}};
	\draw[black, thick] (5.5,1) -- (3.5,1) node [left] {\small{$\ket{2}$}};
	\draw[black, thick] (5.5,1.5) -- (3.5,1.5) node [left] {\small{$\ket{3}$}};
	\draw[black, thick, ->] (4.17,0.5) -- (4.17,0) node [above left] {\footnotesize{$\gamma_{21}$}};
	\draw[black, thick, ->] (4.83,1) -- (4.83,0.5) node [above left] {\footnotesize{$\gamma_{20}$}};
\end{tikzpicture}
	\caption{Depiction of the "problematic" ladder-like structure that renders MAD's non degradable, in the case of $d=4$.}
	\label{fig:ladder-nondeg}
\end{figure}
For any  invertible MAD channel $\Phi_{\Gamma}$ (see \eqref{invertibilitycondition11})
we can check whether it is degradable by observing that the degrading channel $\Lambda _\Gamma$ which maps
${\Phi} _\Gamma$ into its complementary $\tilde{\Phi} _\Gamma$ must fulfil the identity	\begin{equation}
	\label{equation:degradingMAD}
	\Lambda _\Gamma \coloneqq \tilde{\Phi} _\Gamma\circ\Phi ^{-1} _{\Gamma}.
\end{equation}
By construction, this map is linear and trace-preserving, however it is not guaranteed to be completely positive. We can check that numerically with relative ease by computing the eigenvalues of its Choi matrix~\cite{choitheorem,jamiolkowski}.
The problem of finding the complete analitycal degradability conditions on the $\gamma_{ji}$'s is more complicated. The difficulty arises from the fact that the eigenvalue problem involves finding the roots of a polynomial whose degree increases with the dimensionality of the channel. We have found a working recipe that allows us to somewhat simplify the problem and through which the complete degradability region for $4$-dimensional MAD channels was found.

\begin{enumerate}
	\item The first step consists in a heuricstic assumption: if we want the degrading map to be a quantum channel, since this map sends the output state of the laboratory system into the output state of the environment, we would expect density matrix of the latter to have rank not greater than that of the former:
	      \begin{equation}
		      \label{equation:degrankcondition}
		      {\rank \left(\Phi_{\Gamma} (\rho)\right)\geq\rank\left(\tilde{\Phi}_{\Gamma}(\rho)\right)\quad\forall\rho.}
	      \end{equation}
	      As we will see, this condition excludes some valid configurations of degradable MAD's; however, these can be recovered by extending the configurations found under the assumption \eqref{equation:degrankcondition}.
	\item The easiest way to satisfy \eqref{equation:degrankcondition} consists in "turning off" some of the decays: in fact, the dimensionality of the complementary channel's output is equal to the cardinality of the minimal Kraus set of the original channel (see \cite{pipeline}). Therefore, we label all possible $d$-dimensional MAD's with at most $d-1$ non-zero $\gamma_{ji}$'s.
	\item For each of the possible class of MAD's found in the previous step, we compute the eigenvalues of the corresponding degrading maps, resulting in the correct analytical degradability conditions for those classes.
	\item Now we make use of \eqref{equation:qcapacitydeg1} and \eqref{equation:qcapacitydeg2}; starting from the classes found earlier, we may be tempted to "turn on" other decays, purposefully violating \eqref{equation:degrankcondition}. Can we find other degradable configurations by doing so? The answer is yes, as long as the new channel, obtained by turning on the extra decay from the degradable configurations, does not admit decompositions where one of the composing channels is not degradable. We have found that for MAD's, this reduces to excluding all the channels which present the ladder-like decay structure reported in \figref{fig:ladder-nondeg}. In the case of $4$-dimensional MAD's we have found the only possible $4$-decay transition (that violates \eqref{equation:degrankcondition}) which admits degradability under certain conditions for the $\gamma_{ji}$'s.
\end{enumerate}
\section{Quantum capacity  analysis}
\label{section:procedure}
In this section we analyze the quantum capacity $Q$ of
MAD channels.
For antidegradable channels, we already know that $Q=0$. In the case of MAD's once the degradability region is found, in that region we can compute the quantum capacity, the question is then: can we do better? In~\cite{mad3}, the Authors showcased a procedure through which they were able to compute $Q$ for some configurations of $3$-dimensional MAD channels which were neither degradable nor antidegradable. This  can be also applied to higher dimensional channels
by slightly generalizing the procedure along the lines summarized below:
\begin{itemize}
	\item Find degradability conditions as seen in Section \ref{section:degradability}.
	\item Compute quantum capacity in degradable conditions using \eqref{equation:qcapacitydegMAD}.
	\item Starting from a degradability condition, increase the non-zero decay probabilities so that one of the $\gamma_{jj}$'s becomes $0$, then compute the quantum capacity of the resulting channel, mapping the problem to the one outlined in Section~\ref{QforCDMAD}.
	\item If the quantum capacity obtained in the last point is equal to the one at the corresponding border of the degradability condition (e.g. if we computed the capacity at $\gamma_{jj} = 0$, the corresponding border would be the one where the degradability conditions on the $\gamma_{ji}$'s are saturated), we make use of the monotonicity properties for the capacity (see Section \ref{section:monotonicity}) to conclude that the region between the two borders would have the same capacity as the one computed at one of the borders.
	\item We iterate the last three points until possible.
\end{itemize}
In Section \ref{section:example}, we provide an example to clarify the process for  $d=4$ MAD channels. Note that this algorithm exploits computations of the capacity on lower-dimensional MAD's, meaning that the result found e.g. for $4$-dimensional MAD's would also be useful for the capacity computations of  MAD's of larger dimensions.
In the remaining part of the present section we present the two key ingredients of the procedure, namely,
a study of monotonicity properties of the quantum capacity of MAD channels of dimension $d$, and
a detailed analysis of the quantum capacity for MAD channels characterized by complete damping ($\gamma_{jj}=0$) of some levels.
\subsection{Monotonicity properties}
\label{section:monotonicity}
Pipeline inequalities~\cite{pipeline} are important tools to establish a partial  ordering among the quantum capacities of channels.
In our case they imply that if two MAD channels  $\Phi _{\Gamma}$, $\Phi _{\Gamma^{\prime}}$ can be
connected as
\begin{equation}
	\label{equation:monocondition}
	\Lambda _L\circ \Phi _{\Gamma}\circ\Lambda _R =  \Phi _{\Gamma^{\prime}},
\end{equation}
with $\Lambda _L$ and $\Lambda _R$ LCPT channels, then
\begin{eqnarray}
	Q\left(\Phi _{\Gamma^{\prime}}\right) \leq Q\left(\Phi _{\Gamma}\right) \;. \label{pipe3}
\end{eqnarray}
Applying  this result to the
composition rule~\eqref{equation:MADcomposition}, it follows that Eqs.~\eqref{equation:monocondition} and \eqref{pipe3}  hold
in particular whenever  there exist
$\Gamma_L, \Gamma_R\in {\cal M}_d$ transition matrices  such that
\begin{eqnarray}\label{gammaconcatenation}
	\Gamma' = \Gamma_R \Gamma \Gamma_L \;.
\end{eqnarray}
We call ${\cal{S}}(\Gamma)$ the subset of ${\cal M}_d$ whose elements
$\Gamma'$ can be expressed as in \eqref{gammaconcatenation}: they correspond to MAD channels whose quantum capacities cannot be larger than the quantum capacity of $\Phi _{\Gamma}$.
We find also convenient to define
the subsets of ${\cal{S}}(\Gamma)$, ${\cal{S}}_{R}(\Gamma)$ and ${\cal{S}}_{L}(\Gamma)$,
formed  by matrices $\Gamma'$ for which
\eqref{gammaconcatenation} holds  with
$\Gamma_L={\mathds{1}}$ and $\Gamma_R={\mathds{1}}$, respectively.
The convexity of ${\cal M}_d$ implies that both ${\cal{S}}_{R}(\Gamma)$ and ${\cal{S}}_{L}(\Gamma)$ are both convex,  furthermore we can express  ${\cal{S}}(\Gamma)$  as
\begin{equation}
	{\cal{S}}(\Gamma) = \bigcup_{\Gamma' \in {\cal{S}}_{L}(\Gamma)} {\cal{S}}_{R}(\Gamma')=
	\bigcup_{\Gamma' \in {\cal{S}}_{R}(\Gamma)} {\cal{S}}_{L}(\Gamma')\;.
\end{equation}
A compact representation of ${\cal{S}}_{R}(\Gamma)$
can be obtained by describing the
transfer matrices
in terms of collections of $d$-dimensional row vectors,   				\begin{equation}
	\label{equation:definition:rows}
	\Gamma =
	\left(\begin{array}{c}
			\vec{r}_0 \\\hline
			\vec{r}_1 \\\hline
			\vec{r}_2 \\\hline
			\vdots    \\ \hline
			\vec{r}_{d-1}
		\end{array} \right) \quad
	\begin{array}{l}
		\vec{r}_0:=(\;1\;\;,\;0\;,\;0\;,\;0\;,\;\cdots\;, \;0\;)        \\
		\vec{r}_1:=(\gamma_{10},\gamma_{11},\;0\;,\;0\;,\cdots,\;  0\;) \\
		\vec{r}_2
		:=(\gamma_{20},\gamma_{21},\gamma_{22},\;0\;,\cdots,  0\;)      \\
		\quad \vdots                                                    \\
		\vec{r}_d:=(\gamma_{d-1,0},\;\cdots\;,  \gamma_{d-1,d-1})\;,
	\end{array}
\end{equation}
(notice that the $\vec{r}_k$ vector has at most  $k+1$ non zero elements which form a probability distribution).
\begin{lemma}\label{Teorema2:polytope}
	Given $\vec{r}^{\;\prime}_k$ the $k$-th row vector  of
	$\Gamma'\in{\cal{S}}_{R}(\Gamma)$, it can be expressed as a convex combination of the first
	$k$ row vectors of $\Gamma$, i.e.
	there  exists a  probability measure of $k$ elements, $\mathcal{P}_k:=\{ p^{(k)}_{0},
		p^{(k)}_{1}, \cdots, p^{(k)}_{k-1}\}$  such that
	\begin{equation}\label{identitylemma2}
		\vec{r}^{\;\prime}_k = \sum_{\ell=0}^{k} p^{(k)}_{\ell}\vec{r}_{\ell} \;.
	\end{equation}
\end{lemma}
\begin{proof} Since $\Gamma'$ is an element of ${\cal{S}}_{R}(\Gamma)$, there exists
	a transfer matrix $\Gamma_R$ such that $\Gamma'=\Gamma_R \Gamma$. Then \eqref{identitylemma2}
	follows by identifying the entries of $\mathcal{P}_k$ with the first $k$ entries of the row vector of~$\Gamma_R$. \end{proof}

From the above result it follows that ${\cal{S}}_{R}(\Gamma)$ is a convex polytope in ${\cal M}_d$ with at most $d!$ vertexes $\Gamma^{R}_1$, $\Gamma^{R}_2$,  $\cdots$,  that include
the fully damping  transition matrix $\Gamma_{\emptyset}$ 	defined in ~\eqref{equation:definition:transitionmatrixFULL},
$\Gamma$ and the transition matrixes we can obtain from it
by replacing its row vectors with those of with lower indexes in all possible configurations. This construction follows by expressing  the probability set  $\mathcal{P}_k$ as a convex combination of degenerate  distributions
which have only one non zero element. For instance
in the case of $d=2$ we can write
\begin{equation}\label{decompositionprob}
	\mathcal{P}_1=\{ p^{(1)}_{0},
	p^{(1)}_{1}\} =p^{(1)}_{0} \{ 1,
	0\} +p^{(1)}_{1} \{ 0,1\} \;,
\end{equation}
leading to a two vertex polytope with vertexes
\begin{eqnarray}
	\label{equation:definition:rowsconvex}
	\Gamma^{R}_0&=& \left(\begin{array}{c}
			\vec{r}_0 \\\hline
			\vec{r}_0\end{array} \right)=\left(\begin{array}{cc}
			1 & 0 \\				1 & 0\end{array} \right)
	=\Gamma_{\emptyset}\,,\\
	\Gamma^{R}_1 &=&
	\left(\begin{array}{c}
			\vec{r}_0 \\\hline
			\vec{r}_1\end{array} \right) =\left(\begin{array}{cc}
			1 & 0 \\				\gamma_{10} & \gamma_{11}\end{array} \right)=\Gamma \,. \nonumber
\end{eqnarray}
Of course the number of vertexes reduces to one when  $\Gamma^{R}_1=\Gamma^{R}_0$, i.e. when $\gamma_{10}=1$.
The same construction for $d=3$ gives instead a polytope with the following $3!=6$ vertexes
\begin{eqnarray}&&
	\Gamma^{R}_0=
	\left(\begin{array}{c}
			\vec{r}_0 \\\hline
			\vec{r}_0 \\\hline
			\vec{r}_0\end{array} \right)=\Gamma_{\emptyset}\;, \qquad \quad \Gamma^{R}_1 =
	\left(\begin{array}{c}
			\vec{r}_0 \\\hline
			\vec{r}_1 \\\hline
			\vec{r}_2\end{array} \right)=\Gamma,
	\nonumber \\
	&&\Gamma^{R}_2=
	\left(\begin{array}{c}
			\vec{r}_0 \\\hline
			\vec{r}_0 \\\hline
			\vec{r}_2\end{array} \right),\qquad \qquad \quad \Gamma^{R}_3=\nonumber
	\left(\begin{array}{c}
			\vec{r}_0 \\\hline
			\vec{r}_1 \\\hline
			\vec{r}_0\end{array} \right), \\
	\nonumber && \Gamma^{R}_4=
	\left(\begin{array}{c}
			\vec{r}_0 \\\hline
			\vec{r}_0 \\\hline
			\vec{r}_1\end{array} \right),
	\qquad \qquad \quad
	\Gamma^{R}_5=
	\left(\begin{array}{c}
			\vec{r}_0 \\\hline
			\vec{r}_1 \\\hline
			\vec{r}_1\end{array} \right)\;
	\nonumber . \end{eqnarray}
Also in this case the total number of edges can smaller than 6 in special cases where one or more of the above matrices belong on
convex closure of the remaining ones.
\\

Let us now consider the set ${\cal{S}}_{L}(\Gamma)$.
In this case we find if useful to express
transfer matrices
in terms of a $d$-dimensional column vectors, i.e.
\begin{equation}
	\label{equation:definition:columns}
	\Gamma =
	\left(\begin{array}{c|c|c|c|c}
			\vec{c}_0 & \vec{c}_1 & \vec{c}_2 & \;\vdots \; & \vec{c}_{d-1}
		\end{array} \right) \;,
\end{equation}
with
\begin{eqnarray} \nonumber
	{\vec{c}_0}^{\;T}&:=&(1\;,\gamma_{10}\;,\gamma_{20} \;, \cdots\cdots\;,  \gamma_{d-1,0})\;,\\  \nonumber
	\vec{c}_1^{\;T}&:=&(0\;,\gamma_{11}\;,\gamma_{21}\;,\cdots\cdots,  \gamma_{d-1,1})\;,\\
	\nonumber \vec{c}_2^{\;T}
	&:=&(0\;,0\;,\gamma_{22}\;,\gamma_{32}\;,\cdots\;, \gamma_{d-1,2})\;,\\
	&\vdots& \\
	\nonumber \vec{c}_{d-2}^{\;T}&:=&(0\;,\cdots\;,0\;, \gamma_{d-2,d-2}\;, \gamma_{d-1,d-2})\;,\\
	\nonumber \vec{c}_{d-1}^{\;T}&:=&(0\;,0\;,\cdots\cdots\;, 0\;, \gamma_{d-1,d-1})\;.
\end{eqnarray}
Notice that  the vector $\vec{c}_k$ has  at most $d-k$ non zero components. Its elements do not form a probability set, however the sum of  the first $\ell$ vectors has 1's in the first $\ell$ positions,
\begin{eqnarray} \nonumber
	(\vec{c}_0^{\;T}+\vec{c}_1^{\;T})\Big|_{\ell} &=&1\;, \quad \forall \ell =0,1\;,\\
	(\vec{c}_0^{\;T}+\vec{c}_1^{\;T}+\vec{c}_2^{\;T})\Big|_{\ell} &=&1\;, \quad \forall \ell =0,1,2\;,\nonumber\\ &\vdots&\label{normalization}\\  \nonumber
	\left(\sum_{k=0}^{d-1}\vec{c}_k^{\;T}\right)\Big|_{\ell} &=&1\;, \quad \forall \ell =0,1,\cdots, d-1\;.
\end{eqnarray}
\begin{lemma}\label{lemma3}
	Given ${\vec{c}}^{\;\prime}_k$ the $k$-th column vector  of
	$\Gamma'\in{\cal{S}}_{L}(\Gamma)$, it can be expressed as a sum of the
	last $d-k$ column vectors of~$\Gamma$. Specifically
	there  exist  $d$ independent
	probability measures  $\mathcal{P}_0:=\{ p^{(0)}_{0}=1\}$, $\mathcal{P}_1:=\{ p^{(1)}_{0}, p^{(1)}_{1}\}$,
	$\mathcal{P}_2:=\{ p^{(2)}_{0}, p^{(2)}_{1}, p^{(2)}_{2}\}$, $\cdots$, $\mathcal{P}_{d-1}:=\{ p^{(d-1)}_{0}, p^{(d-1)}_{1}, \cdots, p^{(d-1)}_{d-1}\}$,  such that
	\begin{eqnarray}\label{identitylemma3}
		{\vec{c}}^{\;\prime}_k = \sum_{\ell=k}^{d-1} p^{(\ell)}_{k}\vec{c}_{\ell} \;,
	\end{eqnarray}
	for all $k=0,\cdots, d-1$ (notice that the sum over $\ell$ involves elements of
	different measure sets).
\end{lemma}
\begin{proof} Since $\Gamma'$ is an element of ${\cal{S}}_{L}(\Gamma)$, there exists
	a transfer matrix $\Gamma_L$ such that $\Gamma'= \Gamma\Gamma_L$. Then \eqref{identitylemma2}
	follows by identifying the $k$-th probability measure
	$\mathcal{P}_k$   with the first $k$ entries of the $k$-th row vector of $\Gamma_L$.
\end{proof}

We can use the above result  to conclude that  ${\cal{S}}_{L}(\Gamma)$ is also a convex polytope in ${\cal M}_d$ with at most $d!$ vertexes, $\Gamma_1^{L}$, $\Gamma_2^{L}$, $\cdots$ which, among others, include
$\Gamma$ and  $\Gamma_{\emptyset}$.
As for the case of ${\cal{S}}_{R}(\Gamma)$,  this follows by expressing  the $\mathcal{P}_k$'s  as   convex combinations of degenerate  distributions
which have only one non zero element. For instance
for $d=2$, we get that ${\cal{S}}_{L}(\Gamma)$  is the polytope of
vertexes
\begin{eqnarray}\nonumber
	\Gamma^{L}_0 &=&
	\left(\begin{array}{c|c}
			\vec{c}_0 +\vec{c}_1 & 0
		\end{array} \right)=\left(\begin{array}{cc}
			1 & 0 \\				1 & 0\end{array} \right) =\Gamma_{\emptyset}\;,\\
	\Gamma^{L}_1 &=&
	\left(\begin{array}{c|c}
			\vec{c}_0 & \vec{c}_1
		\end{array} \right)=\left(\begin{array}{cc}
			1 & 0 \\				\gamma_{10} & \gamma_{11}\end{array} \right)=\Gamma \;,
	\nonumber
\end{eqnarray}
which incidentally coincides with those of   ${\cal{S}}_{R}(\Gamma)$.
On the contrary for $d=3$
the vertexes of
polytope with the following $3!=6$ vertexes  of ${\cal{S}}_{L}(\Gamma)$ are
\begin{eqnarray*}
	\Gamma^{L}_0 &=&
	\left(\begin{array}{c|c|c}
			\vec{c}_0 +\vec{c}_1+ \vec{c}_2 & 0 & 0
		\end{array} \right)
	=\Gamma_{\emptyset}\;,\\\Gamma^{L}_1
	&=&
	\left(\begin{array}{c|c|c}
			\vec{c}_0 & \vec{c}_1 & \vec{c}_2
		\end{array} \right)=\Gamma \;,
	\\
	\Gamma^{L}_2 &=&
	\left(\begin{array}{c|c|c}
			\vec{c}_0 +\vec{c}_2 & \vec{c}_1 & 0
		\end{array} \right)\;, \\
	\Gamma^{L}_3 &=&
	\left(\begin{array}{c|c|c}
			\vec{c}_0 & \vec{c}_1+\vec{c}_2 & 0
		\end{array} \right)\;, \\\Gamma^{L}_4 &=&
	\left(\begin{array}{c|c|c}
			\vec{c}_0 +\vec{c}_1 & 0 & \vec{c}_2
		\end{array} \right)\;, \\\Gamma^{L}_5 &=&
	\left(\begin{array}{c|c|c}
			\vec{c}_0 +\vec{c}_1 & \vec{c}_2 & 0
		\end{array} \right)\;,
\end{eqnarray*}
where in the last line we invoked~\eqref{normalization}.
Notice that in general these vertexes do not coincide with   those of
${\cal{S}}_{R}(\Gamma)$ for $d=3$, so that ${\cal{S}}_{L}(\Gamma)$ and ${\cal{S}}_{R}(\Gamma)$ are typically different subsets.
\\

A consequence of Lemmas~\ref{Teorema2:polytope},\ref{lemma3} is that
the quantum capacity of MAD channels is monotonic w.r.t. to some
of the
off-diagonal parameters of the matrix $\Gamma$:

\begin{corollary}\label{corollary1}\emph{\bf[Monotonicity under  $\gamma _{10}$]} Let  $\Phi_{\Gamma}$ and $\Phi_{\Gamma'}$ be MAD channels with transitions matrices
	$\Gamma$ of elements $\gamma_{jk}$
	and $\Gamma'$ of elements $\gamma'_{jk}$,
	that only differ on the position
	$(1,0)$  and $(1,1)$,  where
	one has
	\begin{eqnarray}\label{connection10}
		\gamma_{10} &\leq&   \gamma'_{10} \;, \\
		\gamma_{11} &=&  \gamma'_{11} +(\gamma'_{10} -\gamma_{10}) \geq \gamma'_{11}\;, \nonumber
	\end{eqnarray}
	then Eq.~\eqref{pipe3} holds
	(notice that the increase in the off-diagonal term is compensated by a decrease in the associated diagonal element).
\end{corollary}
\begin{proof}
	For $d=3$ this result was proven in~\cite{mad3}. One can generalize this to  arbitrary $d$  observing  that  setting
	\begin{eqnarray}\label{settingpipe}
		{\xi _{1} ^{(0)}}=\frac{\gamma'_{10}-\gamma_{10}}{1-\gamma_{10}}\;,
	\end{eqnarray}
	the probability parameter of the single-decay transition matrix ${\Xi _{1} ^{(0)}}$, one has \begin{equation}\label{firstmono}
		\Gamma' =\Xi _{1} ^{(0)}\; \Gamma \quad
		\Longrightarrow \quad \Gamma'\in {\cal{S}}_{R}(\Gamma)\;.
	\end{equation}
	Accordingly it follows $\Phi_{\Gamma'}=\Phi_{\Gamma} \circ \Phi_{\Xi _{1} ^{(0)}}$ and hence the thesis.
\end{proof}

\begin{corollary}\label{corollary2}\emph{\bf[Monotonicity under  $\gamma _{d-1,k_0}$]} Let  $\Phi_{\Gamma}$ and $\Phi_{\Gamma'}$ be MAD channels with transitions matrices
	$\Gamma$ of elements $\gamma_{jk}$
	and $\Gamma'$ of elements $\gamma'_{jk}$,
	that, given $k_0<d-1$ only differ on the position
	$(d-1,k_0)$ and $(d-1,d-1)$,  where
	one has
	\begin{eqnarray}\label{connection101}
		\gamma_{d-1,k_0} &\leq&   \gamma'_{d-1,k_0} \;, \\
		\gamma_{d-1,d-1} &=&  \gamma'_{d-1,d-1} +(\gamma'_{d-1,k_0} -\gamma_{d-1,k_0})	\nonumber \\
		&\geq& \gamma'_{d-1,d-1}\;, \nonumber
	\end{eqnarray}
	then Eq.~\eqref{pipe3} holds.
\end{corollary}
\begin{proof}
	As in the case of Corrollary~\ref{corollary1}, for $d=3$ this result was proven in~\cite{mad3}. One can generalize this to  arbitrary $d$  observing  that  setting
	\begin{equation}\label{settingpipe2}
		{\xi _{d-1} ^{(k_0)}}=\frac{\gamma'_{d-1,k_0}-\gamma_{d-1,k_0}}
		{\gamma'_{{d-1,d-1}}+(\gamma'_{d-1,k_0}-\gamma_{d-1,k_0})}\;,
	\end{equation}
	the probability parameter of the single-decay transition matrix ${\Xi _{d-1} ^{(k_0)}}$, one has \begin{equation}\label{firstmono1}
		\Gamma' = \Gamma\;  \Xi _{d-1} ^{(k_0)} \quad
		\Longrightarrow \quad \Gamma'\in {\cal{S}}_{L}(\Gamma)\;.
	\end{equation}
	Accordingly it follows $\Phi_{\Gamma'}= \Phi_{\Xi _{d-1} ^{(k_0)}}\circ \Phi_{\Gamma}$ and hence the thesis.
\end{proof}
Notice that for $d=2$ and $d=3$ the above results implies monotonicity
with respect to all the off-diagonal entries of the transition matrix. However,
apart from same special cases
for MAD channel with  $d=4$ reported in Section~\ref{section:example},
we have no evidence of the possibility to extend this result to arbitrary dimensions.

\subsection{Capacity of MAD channels with completely damped levels}
\label{QforCDMAD}

In this section we focus on
MAD channels $\Phi_{\Omega^{(\rm cd)}_{\mathcal{A}}}$ associated with  transfer matrices
$\Omega^{(\rm cd)}_{\mathcal{A}}$ of elements $\omega_{ji}$
which assign complete damping to a collection of levels ${\mathcal{A}}:=\{|k_1\rangle, |k_2\rangle, \cdots, |k_c\rangle\}$, i.e.
\begin{equation}\label{conditionongammaforcdgeneral1}
	{\omega}_{jj}=1-\sum_{i=0}^{j-1}{\omega}_{ji}=0\;, \qquad \forall j\in {\mathcal{A}}\;,
\end{equation}
Notice in particular that from Eq.~(\ref{equation:definition:madchannelkraus1}), it follows that such channels, map all the off-diagonal matrix elements that involve  at least one level that is completely damped into the null operator, i.e.
\begin{equation}
	\label{equation:definition:madchannelcd}
	\Phi_{\Omega^{(\rm cd)}_{\mathcal{A}}} (|j\rangle\langle i|) 	= 0
	\qquad  \forall j\in {\mathcal{A}}, \forall i\neq j\;,
\end{equation}
(the fully damping channel of
\eqref{equation:definition:FULLY} is a special example of these type of transformations).
This property can be used to show that
the quantum capacity of $\Phi_{\Omega^{(\rm cd)}_{\mathcal{A}}}$
involves only maximizations over input states which are not supported on the linear span ${\mathcal{H}_{\mathcal{A}}}:= \mbox{span}[{\mathcal{A}}]$
of the elements of ${\mathcal{A}}$, or
equivalently, that elements of ${\mathcal{H}_{\mathcal{A}}}$ are useless to create optimal communication codes. This can be formalized by introducing the restriction $\Phi_{{\Omega}^{(\rm rs)}_{\bar{\mathcal{A}}}}$ of
$\Phi_{{\Omega}^{(\rm cd)}_{\mathcal{A}}}$   on  the set of linear operators of the
orthogonal complement ${\mathcal{H}_{\bar{\mathcal{A}}}}$
of ${\mathcal{H}_{\mathcal{A}}}$,
\begin{equation}
	\label{equation:restriction}
	\Phi_{{\Omega}^{(\rm rs)}_{\bar{\mathcal{A}}}}(|j\rangle\langle i|) =
	\Phi_{{\Omega}^{(\rm cd)}_{\mathcal{A}}}(|j\rangle\langle i|) \qquad
	\forall j,i\in {\mathcal{H}_{\bar{\mathcal{A}}}}\;. 		\end{equation}
\begin{theorem}\label{Teorema2}
	The quantum capacity of a MAD channel  $\Phi_{\Omega^{(\rm cd)}_{\mathcal{A}}}$ that induces complete damping on all the
	levels of the set ${\mathcal{A}}$, is equal to the capacity of the restricted LCPTP channel
	$\Phi_{{\Omega}^{(\rm rs)}_{\bar{\mathcal{A}}}}$,
	\begin{equation}
		\label{equation:capacityCD}
		Q(\Phi_{\Omega^{(\rm cd)}_{\mathcal{A}}}) = Q(\Phi_{{\Omega}^{(\rm rs)}_{\bar{\mathcal{A}}}})\;.\end{equation}
\end{theorem}
The property~\eqref{equation:capacityCD} was originally proven in~ \cite{mad3} for MAD channels with $d=3$. That proof relies on the fact that the levels involved are either the top one or the first excited one, which greatly simplifies the analysis. As a matter of fact the derivation presented in~ \cite{mad3}
can be easily extended to arbitrary $d$ values, but for the special cases of
sets ${\mathcal{A}}$ which are formed by top levels (say all elements from a given $k$ to $d-1$ included) and/or by the first excited one. To go beyond this we need a new approach that we provide here.
\begin{proof}
	Let's first observe that
	if one were to encode information only on the orthogonal complement ${\mathcal{H}_{\bar{\mathcal{A}}}}$, the channel ${\Phi_{\Omega^{(\rm cd)}_{\mathcal{A}}}}$ would be equivalent to $\Phi_{{\Omega}^{(\rm rs)}_{\bar{\mathcal{A}}}}$; of course, this choice of encoding is not guaranteed to be optimal, therefore:
	\begin{equation}
		\label{equation:completedamping3lb}
		Q(\Phi_{\Omega^{(\rm cd)}_{\mathcal{A}}})\geq
		Q(\Phi_{{\Omega}^{(\rm rs)}_{\bar{\mathcal{A}}}}).
	\end{equation}
	To prove Eq.~\eqref{equation:capacityCD} it is hence sufficient to show that
	$Q(\Phi_{{\Omega}^{(\rm rs)}_{\bar{\mathcal{A}}}})$
	is also greater than or equal to  $Q(\Phi_{\Omega^{(\rm cd)}_{\mathcal{A}}})$.

	An upper bound for the capacity of  $\Phi_{\Omega^{(\rm cd)}_{\mathcal{A}}}$ can be
	obtained using the pipeline inequalities~\eqref{pipe3}
	and a result on the capacity of  Direct Sum (DS) channels~\cite{direct_sum_channels}.
	We recall that  given two Hilbert spaces ${\mathcal{H}}_Z=
			{\mathcal{H}}_X \oplus {\mathcal{H}}_Y$ and
	${\mathcal{H}}_{Z'}=
		{\mathcal{H}}_{X'} \oplus {\mathcal{H}}_{Y'}$ that are  direct sums
	of the orthogonal subspaces
	${\mathcal{H}}_X$, ${\mathcal{H}}_Y$ and  ${\mathcal{H}}_{X'}$, ${\mathcal{H}}_{Y'}$,
	a DS channel  $\Phi^{\rm DS}$ from
	${\mathcal{H}_Z}$ to ${\mathcal{H}}_{Z'}$,
	is a transformation that can be expressed as		direct sum of two individual LCPTP transformations   $\Phi_{XX'}$ and $\Phi_{YY'}$ which  map   ${\mathcal{H}}_X$  into ${\mathcal{H}}_{X'}$ and  ${\mathcal{H}}_Y$ into
	${\mathcal{H}}_{Y'}$  respectively, i.e.
	\begin{equation}\label{generalDS}
		\Phi^{\rm DS}=\Phi_{XX'}\oplus \Phi_{YY'} \;,
	\end{equation}
	hence sending to zero all the off-diagonal operators $\Theta_{XY}$ ($\Theta_{YX}$) which connects
	${\mathcal{H}}_{Y}$ with
	${\mathcal{H}}_{X}$ (resp. ${\mathcal{H}}_{X}$ with
	${\mathcal{H}}_{Y}$), i.e.
	$\Phi^{\rm DS}(\Theta_{XY})=\Phi^{\rm DS}(\Theta_{YX})=0$.
	\begin{figure*}[t]
		\centering
		\includegraphics[width=\textwidth]{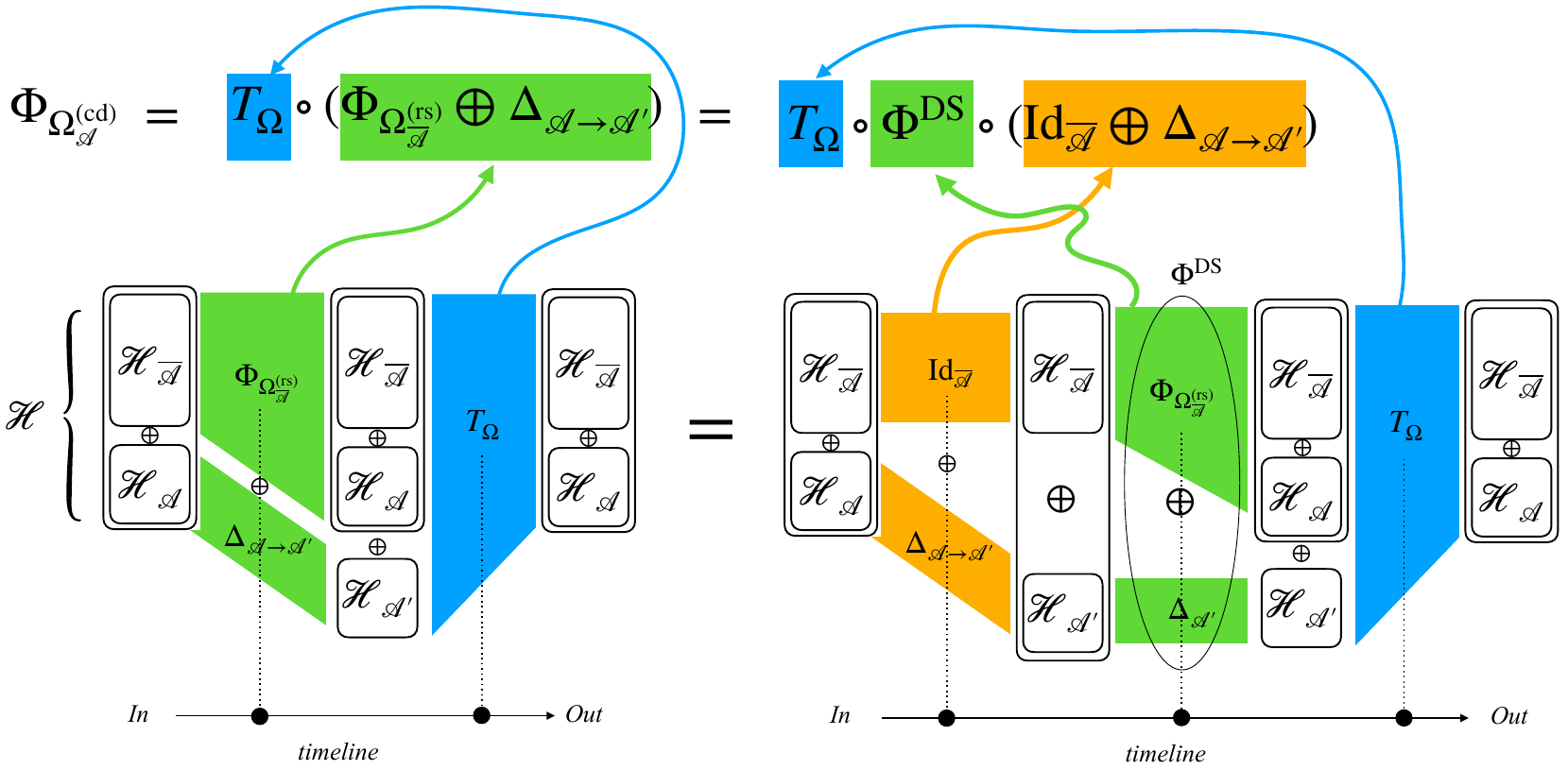}
		\caption{Graphical depiction of the decomposition~\eqref{composition} of the channel
		$\Phi_{{\Omega}^{(\rm cd)}_{{\mathcal{A}}}}$.}
		\label{figureCD}
	\end{figure*}
	The
	DS channel we consider here  is
	\begin{eqnarray}\label{defDS}
		\Phi^{\rm DS}:=
		\Phi_{{\Omega}^{(\rm rs)}_{\bar{\mathcal{A}}}} \oplus \Delta_{\mathcal{A}'}\,,
	\end{eqnarray}
	where we identify $\Phi_{{XX'}}$ with the
	the restricted map $\Phi_{{\Omega}^{(\rm rs)}_{\bar{\mathcal{A}}}}$
	defined in Eq.~\eqref{equation:restriction}, and
	$\Phi_{{YY'}}$ with
	the	map
	$\Delta_{\mathcal{A}'}$ that induce complete dephasing among the element of
	the compulational basis of ${\mathcal{H}_{\mathcal{A}'}}$, i.e.
	\begin{equation}\label{DeltaAprime}
		{\Delta}_{{\mathcal{A}'}}(|i\rangle_{\mathcal{A}'} \langle
		j|) = \delta_{i,j}\, |j \rangle_{\mathcal{A}'} \langle
		j| \quad \forall i,j \in {\mathcal{A}'}\;.
	\end{equation}
	In this scenario one can prove that the quantum capacity
	of $\Phi^{\rm DS}$ reduces to the quantum capacity of
	$\Phi_{{\Omega}^{(\rm rs)}_{\bar{\mathcal{A}}}}$ alone, i.e.
	\begin{equation}
		\label{equation:dscapacity1}
		Q(\Phi^{\rm DS})=Q(\Phi_{{\Omega}^{(\rm rs)}_{\bar{\mathcal{A}}}})\;.
	\end{equation}
	This follows from the fact
	that the  coherent information of a DS channel
	on a generic input state $\rho\in{\cal H}_Z$  corresponds to the maximum of the coherent information of products of  the individual components $\Phi_{XX'}$ and $\Phi_{YY'}$~\cite{direct_sum_channels}, and from the fact that for the special choice~\eqref{DeltaAprime} the contributions associated with the terms $\Phi_{YY'}$  can be neglected since this are entanglement breaking maps~\cite{holevo_book}, see Appendix~\ref{APPDS} for details.
	We now observe that the channel $\Phi_{{\Omega}^{(\rm cd)}_{{\mathcal{A}}}}$
	can be expressed as a concatenation of LPCPT maps that includes $\Phi^{\rm DS}$. Specifically, define
	${\Delta}_{{\mathcal{A}}\rightarrow{\mathcal{A}'}}$ the map which first moves ${\mathcal{H}_{\mathcal{A}}}$ into ${\mathcal{H}_{\mathcal{A}'}}$ and then acts with $\Delta_{\mathcal{A}'}$, i.e.
	\begin{equation}
		{\Delta}_{{\mathcal{A}}\rightarrow{\mathcal{A}'}}(|i\rangle_{\mathcal{A}} \langle
		j|) = \delta_{i,j}\, |j \rangle_{\mathcal{A}'} \langle
		j| \quad \forall i,j \in {\mathcal{A}}\;.
	\end{equation}
	Define also ${T}_{\Omega}$ the map which connects ${\mathcal{H}}\oplus {\mathcal{H}_{\mathcal{A}'}}$
	to ${\mathcal{H}}$ by acting as the identity on ${\mathcal{H}}$, and as a decoherence channel on
	${\mathcal{H}_{\mathcal{A}'}}$ which reverses the populations of its computational levels to
	into mixture of  levels  of ${\mathcal{H}}$ with transitions probabilities given by the elements~\eqref{conditionongammaforcdgeneral1} of the transition matrix $\Omega^{(\rm cd)}_{\mathcal{A}}$, i.e.
	\begin{eqnarray}{T}_{\Omega}(|j\rangle \langle
		i|) =\left\{\begin{array}{ll}|j\rangle \langle
			i| \quad \qquad  \qquad\forall |j\rangle, |i\rangle\in {\mathcal{H}} \;, \\ \\
			\delta_{i,j}\sum_{i'=0}^{j-1} \omega_{ii'} |i' \rangle \langle
			i'|                                                                      \\
			\qquad \qquad \qquad \forall |j\rangle, |i\rangle\in {\mathcal{H}_{\mathcal{A}'}}
			\;,\nonumber                                                             \\ \\
			0\quad\qquad \forall |j\rangle   \in {\mathcal{H}_{\mathcal{A}'}}\;,
			\forall |i\rangle \in {\mathcal{H}}
			\;.\end{array} \right. \nonumber
	\end{eqnarray}
	It hence follows that we can write
	\begin{eqnarray}	\nonumber
		&&\Phi_{{\Omega}^{(\rm cd)}_{{\mathcal{A}}}}=
		T_{\Omega} \circ
		(\Phi_{{\Omega}^{(\rm rs)}_{\bar{\mathcal{A}}}} \oplus {\Delta}_{{\mathcal{A}}\rightarrow{\mathcal{A}'}} )
		\\&&\; = \nonumber T_{\Omega} \circ (
		\Phi_{{\Omega}^{(\rm rs)}_{\bar{\mathcal{A}}}} \oplus (\Delta_{\mathcal{A}'} \circ {\Delta}_{{\mathcal{A}}\rightarrow{\mathcal{A}'}}
		))
		\\&&\; = \nonumber  T_{\Omega}\circ
		(\Phi_{{\Omega}^{(\rm rs)}_{\bar{\mathcal{A}}}} \oplus \Delta_{\mathcal{A}'}) \circ ({\rm Id}_{\bar{\mathcal{A}}} \oplus
		{\Delta}_{{\mathcal{A}}\rightarrow{\mathcal{A}'}})
		\\
		&&\; ={T}_{\Omega} \circ \Phi^{\rm DS}\circ
		({\rm Id}_{\bar{\mathcal{A}}} \oplus
		{\Delta}_{{\mathcal{A}}\rightarrow{\mathcal{A}'}})
		\,,\label{derivation}
		\label{composition}
	\end{eqnarray}
	where  ${\rm Id}_{\bar{\mathcal{A}}}$ is the identity map on ${\mathcal{H}_{\bar{\mathcal{A}}}}$, and where we used \eqref{defDS} and the fact that ${\Delta}_{{\mathcal{A}}\rightarrow{\mathcal{A}'}}= \Delta_{\mathcal{A}'} \circ {\Delta}_{{\mathcal{A}}\rightarrow{\mathcal{A}'}}$, see Fig.~\ref{figureCD}.
	The last identity in Eq.~(\ref{derivation}) is the identity we are looking for: invoking the pipeline inequality it allows us to reverse the inequality in \eqref{equation:completedamping3lb},
	\begin{equation}
		\label{equation:completedamping3lbref}
		Q(\Phi_{\Omega^{(\rm cd)}_{\mathcal{A}}})\leq
		Q(\Phi_{{\Omega}^{(\rm rs)}_{\bar{\mathcal{A}}}})\;,
	\end{equation}
	and hence to prove Eq.~\eqref{equation:capacityCD}.
\end{proof}
From Theorem~\ref{theorem1} the following implication can be easily derived:
\begin{remark}\label{remark1} In the special case where the  MAD channel $\Phi_{\Omega^{(\rm cd)}_{\mathcal{A}}}$  acts as the identity on ${\mathcal{H}_{\bar{\mathcal{A}}}}$,
	Eq.~\eqref{equation:capacityCD} gives us
	\begin{equation}
		\label{equation:capacityCDspecial}
		Q(\Phi_{\Omega^{(\rm cd)}_{\mathcal{A}}}) =\log_2(d_{\bar{\mathcal{A}}}) = \log_2(d-d_{\mathcal{A}})
		\;,\end{equation}
	where $d_{\bar{\mathcal{A}}}$ and $d_{{\mathcal{A}}}=d-d_{\bar{\mathcal{A}}}$ are the dimension of ${\mathcal{H}_{\bar{\mathcal{A}}}}$
	and ${\mathcal{H}_{\mathcal{A}}}$ respectively.
\end{remark}

\begin{remark}\label{remark2}
	If no transitions repopulate the levels of the completely damped subset
	$\mathcal{A}$,  i.e. if  the matrix ${\Omega}^{(\rm cd)}_{{\mathcal{A}}}$ does not map elements of
	$\bar{\mathcal{A}}$ into $\mathcal{A}$,
	the restricted channel $\Phi_{{\Omega}^{(\rm rs)}_{\bar{\mathcal{A}}}}$ is formally equivalent to a MAD channel of dimension $d_{\bar{\mathcal{A}}}$ with transition matrix
	$\Gamma^{(\rm rs)}$  given by the sub-matrix of ${\Omega}^{(\rm cd)}_{{\mathcal{A}}}$ on~$\mathcal{A}$. For instance, if $\Phi_{{\Omega}^{(\rm cd)}_{{\mathcal{A}}}}$ is a 4-dimensional MAD channel and $d_{{\mathcal{A}}}=1$, its capacity can be computed as the capacity of a 3-dimensional
	MAD channel operating on the levels that are not affected by complete decay.
\end{remark}

\section{Quantum capacity of $d=4$ MAD channels }
\label{section:example}
From the result of Section~\ref{section:monotonicity} we know that the quantum capacity of a $4$ dimensional MAD channel is monotonous w.r.t. to the off-diagonal elements $\gamma_{10}$, $\gamma_{30}$, $\gamma_{31}$, and $\gamma_{32}$. Yet it
is unclear whether these channels present monotonous capacities under the transition probabilities $\gamma _{20}$ and $\gamma_{21}$. A partial answer to this problem can be obtained when additional assumptions are made. Suppose for instance that $\Phi_{\Gamma}$ is invertible and that an extra channel
$\Phi_{\Gamma'}$ is connected as in~\eqref{equation:monocondition}   with
either one of $\Lambda _L$ or $\Lambda _R$ being the identity map.  Then, utilizing the inverse map \eqref{equation:definition:MADinverse} we can write:
\begin{eqnarray} \left\{ \begin{array}{ll}
		\label{equation:monochoiL}
		\Lambda _L = \Phi _{\Gamma^{\prime}}\circ\Phi ^{(-1)} _{\Gamma} &
		\quad \mbox{for $\Lambda _R=\mathrm{Id}$},                                                                   \\ \\
		\Lambda _R = \Phi ^{(-1)} _{\Gamma}\circ\Phi _{\Gamma^{\prime}}
		                                                                & \quad \mbox{for $\Lambda _L=\mathrm{Id}$}.\end{array}\right.
\end{eqnarray}
Note that, by construction, both $\Phi _{\Gamma^{\prime}}\circ\Phi ^{(-1)} _{\Gamma} $ and
$\Phi ^{(-1)} _{\Gamma}\circ\Phi _{\Gamma^{\prime}}$
are linear and trace preserving; verifying their complete positiveness would imply that they are quantum channels, which can be done by verifying the positive semi-definiteness of their Choi matrices~\cite{choitheorem,jamiolkowski}. When this holds we can invoke Eq.~\ref{pipe3} to conclude that
$Q(\Phi_{\Gamma})$ is larger than or equal to  $Q(\Phi_{\Gamma'})$.
\\

\textbf{Monotonicity under $\gamma _{21}$:}
Assume that the transition matrices $\Gamma$ and
$\Gamma^{\prime}$ have identical entries but for the elements in position
$(2,1)$ and $(2,2)$ where we have
\begin{eqnarray}\label{connection21}
	\gamma_{21} &\leq&   \gamma'_{21} \;, \\
	\gamma_{22} &=&  \gamma'_{22} +(\gamma'_{21} -\gamma_{21}) \geq \gamma'_{22}\;. \nonumber
\end{eqnarray}
Computing the eigenvalues of the Choi matrices of $\Lambda _L,\Lambda _R$ in \eqref{equation:monochoiL} one finds that the former is only positive semi-definite if:
\begin{equation}
	\label{equation:gamma21mono1}
	\gamma_{32}=0,
\end{equation}
while the latter is positive semi-definite under the following condition:
\begin{equation}
	\label{equation:gamma21mono}
	\gamma_{20}-\gamma_{10}\geq 0,
\end{equation}
Accordingly the capacity functionals of a $4$-dimensional MAD channel are monotonous under $\gamma _{21}$ if $\gamma_{32} = 0$ or $\gamma_{20}\geq\gamma_{10}$.
\\

\textbf{Monotonicity under   $\gamma_{20}$}:
Assume that the transition matrices $\Gamma$ and
$\Gamma^{\prime}$ have identical entries but for the elements in position
$(2,0)$ and $(2,2)$ where we have
\begin{eqnarray}\label{connection21_1}
	\gamma_{20} &\leq&   \gamma'_{20} \;, \\
	\gamma_{22} &=&  \gamma'_{22} +(\gamma'_{20} -\gamma_{20}) \geq \gamma'_{22}\;. \nonumber
\end{eqnarray}
In this case the eigenvalues of the Choi matrices for $\Lambda _L$ and $\Lambda _R$ in~\eqref{equation:monochoiL}
one finds that the former is positive semi-definite only if:
\begin{equation}
	\label{equation:gamma21mono2}
	\gamma_{32}=0,
\end{equation}
while the latter is positive semi-definite under the following condition:
\begin{equation}
	\label{equation:gamma20mono}
	1-\gamma_{21}-\gamma_{10}\geq 0,
\end{equation}
Accordingly the capacity functionals of a $4$-dimensional MAD channel are monotonous under under $\gamma _{20}$ if $\gamma_{32} = 0$ or $\gamma_{21}\leq 1-\gamma_{10}$.

\subsection{Example}\label{sec:example}

\begin{figure}
	\centering
	\begin{subfigure}[b]{.4\linewidth}
		\centering\includegraphics[width=.9\linewidth]{figures/example/3Bschemecomplete3.tex}

		\vspace{.25\linewidth}
		\centering\includegraphics[width=.9\linewidth]{figures/example/3Bschemealtcomplete.tex}
	\end{subfigure}
	\hspace{.05\linewidth}
	\centering
	\begin{subfigure}[b]{.5\linewidth}
		\centering\includegraphics[width=.9\linewidth]{figures/example/3Btotal.tex}
	\end{subfigure}
	\caption{On the left we have depictions of the channel described by \eqref{GAMMA} only has $3$ possible decay routes. The top one is the "original" channel, the bottom one is the one obtained by applying level permutations, described by \eqref{GAMMAprime}. On the right, we have the parameter space of the channel; we colored it to differentiate between the different regions of interest: The degradability region $D$ is in blue, the $ND1$ region is in purple, the $ND2$ region is in yellow and the $ND3$ region is in green}
	\label{fig:example:def}
\end{figure}

\begin{figure}
	\centering\includegraphics[width=.9\linewidth]{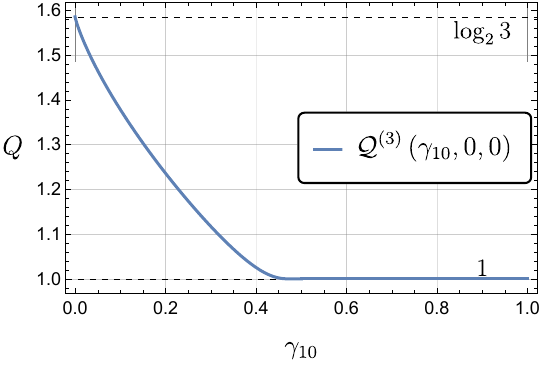}
	\caption{Plot of the function ${\cal Q}^{(3)}(\gamma_{10},0,0)$}
	\label{fig:example:adc3}
\end{figure}

Let us now focus on a special case of $d=4$ MAD channel
depicted in \figref{fig:example:def} together with its parameter space and the degradability region. This
transformation is characterized by only three non-zero off-diagonal elements, $\gamma _{10}$, $\gamma _{30}$
and $\gamma _{32}$:
\begin{equation}
	\label{GAMMA}
	\Gamma = \left(\begin{array}{cccc}
			1             & 0           & 0           & 0           \\
			{\gamma_{10}} & \gamma_{11} & 0           & 0           \\
			0             & 0           & 1           & 0           \\
			{\gamma_{30}} & 0           & \gamma_{32} & \gamma_{33}\end{array} \right)\;,
\end{equation}
where of course $\gamma_{11}=1-\gamma_{10}$,
$\gamma_{33}=1-\gamma_{30}-\gamma_{32}$.
Since condition~\eqref{equation:antidegconditions}
cannot be met,  Theorem~\ref{theorem1} implies that the channel
$\Phi_{\Gamma}$ has always non zero capacity. As a matter of fact,
since the restriction of the channel to the  subspace spanned by vectors
$|0\rangle$ and $|2\rangle$, is the identity map, so that one has that
$Q(\Phi_{\Gamma})$ is always larger than or equal to the quantum capacity of the identity qubit channel, i.e. 1.
An even more stringent lower bound
\begin{equation}\label{result0}
	Q(\Phi_{\Gamma})\geq {\cal Q}^{(3)}(\gamma_{10})\geq 1\;,
\end{equation}
can be obtained by restricting the encoding in the subspace spanned by the the first three levels.
Under this condition the map reduces to the
the 3-dimensional  MAD channel whose
$3\times 3$
transition matrix $\Gamma^{(\rm rs)}$ is formed by
the first three rows and columns of $\Gamma$: its value
${\cal Q}^{(3)}(\gamma_{10},0,0)$ is a monotonically decreasing function of   $\gamma_{10}$ which was computed in~\cite{mad3} and reported here in \figref{fig:example:adc3}.

The  degradable region $D$ for transition matrix~$\Gamma$  of the form~\eqref{GAMMA}
is  identified by the inequalities
\begin{equation}\left\{ \begin{array}{l}
		\frac{1}{2}\leq \gamma_{11}\leq 1 , \\
		\frac{1}{2}\leq \gamma_{33}\leq 1 ,
	\end{array}	\right. \Longleftrightarrow
	\left\{ \begin{array}{l}
		\gamma_{10}\leq \frac{1}{2}, \\
		\gamma_{30}+\gamma_{32}\leq \frac{1}{2},
	\end{array}	\right.
\end{equation}
and represented by the blue volume in~\figref{fig:example:def}:  here we can compute $Q(\Phi_{\Gamma})$ using \eqref{equation:qcapacitydegMAD}. To  extend this computation on the remaining part of the parameter space, we adopt the procedure detailed at the beginning of Section~\ref{section:procedure}.

Consider first the non-degradability region $ND1$ identified by the inequalities
\begin{equation}\left\{ \begin{array}{l}
		\frac{1}{2}\leq \gamma_{11}\leq 1 , \\
		\gamma_{33}\leq 	\frac{1}{2} ,
	\end{array}	\right. \Longleftrightarrow
	\left\{ \begin{array}{l}
		\gamma_{10}\leq \frac{1}{2}, \\
		\gamma_{30}+\gamma_{32}\geq \frac{1}{2},
	\end{array}	\right.\label{ND1}
\end{equation}
which corresponds to the purple volume in \figref{fig:example:def}. Let  focus on the borders of $ND1$ identified by the planes $\gamma_{33} = 0$ (i.e. $\gamma_{30}+\gamma_{32} = 1$) and $\gamma_{33} = 1/2$ (i.e. $\gamma_{30}+\gamma_{32} = 1/2$).
The first plane is formed by transition matrices where level~$|3\rangle$ is completely damped:  here  we can invoke Theorem~\ref{Teorema2}
and compute the corresponding quantum capacity under the constraint  that input states have zero population on level~$|3\rangle$. Furthermore, thanks to Remark~\ref{remark2},
in this case the capacity of the map is equal to the capacity
${\cal Q}^{(3)}(\gamma_{10},0,0)$  of 3-dimensional  MAD channel with the $3\times 3$
transition matrix  $\Gamma^{(\rm rs)}$  defined above.
The second plane separates $ND1$ and the degradability region $D$ allowing us to use~\eqref{equation:qcapacitydegMAD}  to compute the capacity.
By numerical analysis we observe that, for fixed $\gamma_{10}\leq 1/2$,
this value is also provided by the function
${\cal Q}^{(3)}(\gamma_{10})$, allowing us to conclude that
$Q(\Phi_{\Gamma})$ take identical values on two planes:
\begin{eqnarray}\label{result1}
	Q(\Phi_{\Gamma})\Big|_{\gamma_{33}=\frac{1}{2},\gamma_{10}} &=& Q(\Phi_{\Gamma})\Big|_{\gamma_{33}=0,\gamma_{10}} \\\nonumber
	&=& {\cal Q}^{(3)}(\gamma_{10},0,0)\;.
\end{eqnarray}
It is worth stressing that while by construction $Q(\Phi_{\Gamma})\Big|_{\gamma_{33}=0,\gamma_{10}}$  explicitly carries no
dependence upon the parameters $\gamma_{30}$ and $\gamma_{32}$,   in principle $Q(\Phi_{\Gamma})\Big|_{\gamma_{33}=\frac{1}{2},\gamma_{10}}$ may exhibit such dependence:
the numerical evaluation that led us to ~\eqref{result1} however does not reveal it, implying that the capacity can be achieved without populating level~$|3\rangle$.
Most importantly exploiting
the monotonicity w.r.t. $\gamma_{32}$ and $\gamma_{33}$ outlined in Corollary~\ref{corollary2}, we can conclude that the same property
apply for all points inside $ND1$, so that
the quantum
capacity $Q(\Phi_{\Gamma})$ must be constant for fixed $\gamma_{10}$ and equal to the value indicated in~\eqref{result1},~i.e.
\begin{equation}\label{result11}
	Q(\Phi_{\Gamma})\Big|_{ND1} = {\cal Q}^{(3)}(\gamma_{10},0,0)\;.
\end{equation}
We can repeat the same process in order to extend the computation of the quantum capacity to the non-degradable region $ND2$ defined by the inequalities
\begin{equation}\left\{ \begin{array}{l}
		\gamma_{11}\leq \frac{1}{2} , \\
		\frac{1}{2}\leq  \gamma_{33}\leq 	1 ,
	\end{array}	\right. \Longleftrightarrow
	\left\{ \begin{array}{l}
		\gamma_{10}\geq \frac{1}{2}, \\
		\gamma_{30}+\gamma_{32}\leq \frac{1}{2},
	\end{array}	\right.\label{ND2}
\end{equation}
corresponding to the yellow volume in \figref{fig:example:def}.
In this case however we first  employ some level-swap unitaries~\eqref{equation:unitaryconjugation:gamma}  to obtain a unitarily covariant channel with the same capacities as the original, as depicted on the bottom left of \figref{fig:example:def}, leading to a new transition matrix:
\begin{equation}
	\label{GAMMAprime}
	\Gamma' = \left(\begin{array}{cccc}
			1             & 0           & 0           & 0           \\
			0             & 1           & 0           & 0           \\
			\gamma_{30}   & \gamma_{32} & \gamma_{33} & 0           \\
			{\gamma_{10}} & 0           & 0           & \gamma_{11}\end{array} \right)\;,
\end{equation}
The reasoning that follows is the same as before; thanks to Remark~\ref{remark2}, the quantum capacity at the border $\gamma_{11} = 0$ is the same of that of a $3$-dimensional MAD described by the matrix $\Gamma^{\prime(\rm rs)}$, formed by
the first three rows and columns of $\Gamma'$:

\begin{equation}
	Q(\Phi_{\Gamma})\Big|_{\gamma_{11}=0,\gamma_{30},\gamma_{32}} = {\cal Q}^{(3)}(0,\gamma_{30},\gamma_{32})\;.
\end{equation}
The right-hand side of this equation was already computed in \cite{mad3}. We numerically compare this value to that of $Q(\Phi_{\Gamma})\Big|_{\gamma_{11}=\frac{1}{2},\gamma_{30},\gamma_{32}}$ and we find them to be the same, leading to:
\begin{eqnarray}\label{equation:result2}
	Q(\Phi_{\Gamma})\Big|_{\gamma_{11}=\frac{1}{2},\gamma_{30},\gamma_{32}} &=& Q(\Phi_{\Gamma})\Big|_{\gamma_{11}=0,\gamma_{30},\gamma_{32}} \\\nonumber
	&=& {\cal Q}^{(3)}(0,\gamma_{30},\gamma_{32})\;,
\end{eqnarray}
which, thanks to Corollary~\ref{corollary2}, leads to:
\begin{equation}\label{result22}
	Q(\Phi_{\Gamma})\Big|_{ND2} = {\cal Q}^{(3)}(0,\gamma_{30},\gamma_{32})\;.
\end{equation}


Finally, we can compute the quantum capacity in the non-degradability region $ND3$ depicted in green in \figref{fig:example:def} and identified by the conditions
\begin{equation}\left\{ \begin{array}{l}
		\gamma_{11}\leq \frac{1}{2} , \\
		\gamma_{33}\leq 	\frac{1}{2} ,
	\end{array}	\right. \Longleftrightarrow
	\left\{ \begin{array}{l}
		\gamma_{10}\geq \frac{1}{2}, \\
		\gamma_{30}+\gamma_{32}\geq \frac{1}{2}.
	\end{array}	\right.\label{ND3}
\end{equation}
From ~\ref{result0} and the monotonicity property w.r.t.  $\gamma_{10}$ (Corollary~\ref{corollary1}) we have that for fixed $\gamma_{30}$ and $\gamma_{32}$, the following inequalities hold
\begin{eqnarray}\nonumber
	\!\!\!\!\!\!Q(\Phi_{\Gamma})\Big|_{ND3,\gamma_{10}=1/2}	 &\geq& Q(\Phi_{\Gamma})\Big|_{ND3,\gamma_{10}} \\ &\geq& {\cal Q}^{(3)}(\gamma_{10})\geq 1\;.\label{resultND3}
\end{eqnarray}
Notice next that $Q(\Phi_{\Gamma})\Big|_{ND3,\gamma_{10}=1/2}$
is the capacity value  on the plane that separates $ND3$ from $ND1$ which, according to
Eq.~\eqref{result11},  corresponds to ${\cal Q}(\gamma_{10}=1/2)=1$. Now all other points of the volume
can be obtained from those of such plane, increasing $\gamma_{10}$
from $1/2$ to $1$. We can hence conclude that
\begin{equation}\nonumber
	Q(\Phi_{\Gamma})\Big|_{ND3}	 = 1\;,\label{resultND3fin}
\end{equation}
implying that as for $ND1$ also in this case the optimal encoding can be achieved without populating level $\ket{3}$. We can repeat the same reasoning starting from the plane separating $ND2$ and $ND3$, arriving at the same result, meaning that also level $\ket{1}$ should not be populated for the optimal encoding, as is expected since encoding only in levels $\ket{0},\ket{2}$ leads to a $2$-dimensional identity channel, which has capacity $Q(\mathrm{Id} _2) = 1$.

\section{A conjecture}
\label{subsection:conjecture}
The antidegradability conditions \eqref{equation:antidegconditions} provide a powerful tool for identifying "useless" channels; the level-by-level structure of the conditions tempt us to take the result one step further: what if each time a single level $j$ degrades following $\gamma_{j0}-\gamma_{jj}\geq 0$, it means that the $j$ level needs to be excluded in order to achieve optimal encoding? That is:
\begin{equation}
	\label{equation:conjecture}
	\exists j: \gamma_{j0} - \gamma_{jj}\geq 0 \overset{?}{\Rightarrow} | j\rangle \nsubseteq \textrm{ optimal encoding}.
\end{equation}
If this were true, it would mean that the quantum capacity of a $d$-dimensional MAD channel satisfying $\gamma_{j0}  -\gamma_{jj}\leq 0$ for some $j$ would be equal to that of a different channel which takes a $(d-1)$-dimensional input system and outputs a $d$-dimensional system, and this capacity might actually be computable. Theorem~\ref{Teorema2} reveals that is the case when $j$-th level undergoes to complete damping, i.e. when $\gamma_{jj} =0$. More generally we can verify the validity of \eqref{equation:conjecture} in all the cases where we were able to actually compute the quantum capacity using the procedures illustrated in previous sections. For instance for $d=4$ the conjecture holds for subregions of the non degradability region $ND3$ and $ND1$ of the channel analyzed in Sec.~\ref{sec:example} where the condition $\gamma_{j0} - \gamma_{jj}\geq 0$ is verified with $j=3$, and for the regions $ND2$ and $ND3$ where instead it applies to $j=1$.
The same result applies also for the  $d=3$ MAD channels solved in~\cite{mad3}.
To strengthen the argument here go back to these and extend the results analyzed there to include new regions of the parameter space which were left unsolved by previous works. As we shall see, also for them the conjecture is correct.

\subsection{Back to MAD3}
\label{subsubsection:mad3}

\begin{figure}
	\centering
	\begin{subfigure}[b]{0.45\textwidth}
		\centering
		\includegraphics[width=1.0\textwidth]{figures/parameterSpace.tex}
		\caption{Parameter space of $3$-dimensional MAD channels. We want to confirm the conjecture \eqref{equation:conjecture} in the blue region enclosed in the ABCEFG vertices.}
		\label{fig:MAD3-parameterSpace}
	\end{subfigure}
	\hfill
	\begin{subfigure}[b]{0.45\textwidth}
		\centering
		\resizebox{1.0\textwidth}{!}{	\begin{tikzpicture}
		\definecolor{mainpurple}{RGB}{128, 92, 156} 
		\draw[->] (-1,0) -- (8,0) node[right] {$\gamma_{21} = 1-2\gamma_{20}$};
		\draw[->] (0,-1) -- (0,6) node[above] {$\gamma_{10}$};
		
		\draw (0,1)--(-.1,1) node[left] {$0$};
		\draw (0,5)--(-.1,5) node[left] {$\frac{1}{2}$};
		
		\draw (1,0)--(1,-.1) node[below] {$0$};
		\draw (4,0)--(4,-.1) node[below] {$\frac{1}{2}$};
		\draw (7,0)--(7,-.1) node[below] {$1$};
		
		\draw[thick, mainpurple] (1,1) rectangle (7,5);
		
		\node[below left, mainpurple] at (1,1) {C};
		\node[below right, mainpurple] at (7,1) {A};
		\node[above left, mainpurple] at (1,5) {G};
		\node[above right, mainpurple] at (7,5) {E};
		
		\draw[decoration={brace,mirror,raise=5pt},decorate, blue] (1,1) -- node[below=6pt] {${\cal Q}^{(2)}(0)=1$} (7,1);
		\draw[decoration={brace,mirror,raise=5pt},decorate, blue] (1,1.1) -- node[right=6pt] {${\cal Q}^{(2)}(\gamma_{10})$} (1,4.9);
		\draw[decoration={brace,mirror,raise=5pt},decorate, blue] (7,1.1) -- node[right=6pt] {${\cal Q}^{(2)}(\gamma_{10})$} (7,4.9);
		\draw[decoration={brace,raise=5pt},decorate, blue] (1,5) -- node[above=6pt] {${\cal Q}^{(2)}(1/2)=0$} (7,5);

	\end{tikzpicture}}
		\caption{Depiction of the ACGE rectangle of \figref{fig:MAD3-parameterSpace}. We know the capacity along its edges, we hypothesize that it will remain equal to ${\cal Q}^{(2)}(\gamma_{10})$ inside the rectangle.}
		\label{fig:acge}
	\end{subfigure}
	\caption{Top figure is a depiction of the parameter space of $3$-dimesional MAD's, bottom figure represents a region of that parameter space.}
\end{figure}
A $d=3$ MAD channel is fully determined by only three transition probabilities, i.e. $\gamma_{10}$, $\gamma_{20}$, and $\gamma_{21}$, which define the 3D polytope
\begin{equation}\left\{ \begin{array}{l}
		0\leq \gamma_{10}, \gamma_{20},\gamma_{21}\leq 1 , \\
		0\leq  \gamma_{20}+ \gamma_{21}\leq 	1 ,
	\end{array}	\right. \label{MAD3}
\end{equation}
as the physical  parameter region of the problem.
Here we compute the quantum capacity in the cheese-wedge region R  of vertexes ABCDEF  which was not covered in
Ref.~\cite{mad3}. Such region, depicted in~\figref{fig:MAD3-parameterSpace}, is formally identified by the conditions
\begin{equation}\left\{ \begin{array}{l}
		0\leq \gamma_{10}\leq \frac{1}{2} , \\
		\gamma_{20}+ \gamma_{21}\leq 	1 ,
		\\
		1 \leq  2\gamma_{20}+ \gamma_{21} .
	\end{array}	\right. \label{RRR}
\end{equation}
Notice that for all these points we have
\begin{equation} \gamma_{20} - \gamma_{22}\geq 0\;,\label{conjexample}
\end{equation}
so that, according to the conjecture~\ref{equation:conjecture} we expect that the capacity value should be obtained by only using encoding that involve
the level $|0\rangle$ and $|1\rangle$. This formally maps the problem into a qubit
damping channel of probability amplitude $\gamma_{10}$ whose
capacity ${\cal Q}^{(2)}(\gamma_{10})$ was derived in~\cite{adc}:
\begin{equation}
	\label{equation:mad3conjecture}
	Q(R) ={\cal Q}^{(2)}(\gamma_{10}).
\end{equation}
To prove that this is indeed the case we start observing that
on the face ABFE of R (identified by the condition $\gamma_{20}+ \gamma_{21}=1$, i.e.
$\gamma_{22}=0$), Eq.~\eqref{equation:mad3conjecture} was
derived~\cite{mad3} (it can also be derived as a consequence of Theorem~\ref{Teorema2}). We next focus on the face
ACGE of R defined by the condition $2\gamma_{20}+ \gamma_{21}=1$.
At the borders of such rectangle (i.e. on the sides
$AC$, $CG$, $GE$, and $EA$) the capacity was computed in Ref.~\cite{mad3}
with values which are also given by~\eqref{equation:mad3conjecture}, specifically
\begin{eqnarray}\nonumber \label{equation:mad3values}
	Q(AC) &=& Q(ACGE) \Big|_{\gamma_{10}=0} \!\!\!\!\!\!= {\cal Q}^{(2)}(\gamma_{10}=0) =1\;, \\ \nonumber
	Q(CG) &=& Q(ACGE) \Big|_{\gamma_{20}=\frac{1}{2}} \!\!\!\!\!\!= {\cal Q}^{(2)}(\gamma_{10})\;, \\ \nonumber
	Q(GE) &=& Q(ACGE)
	\Big|_{\gamma_{10}=\frac{1}{2}} \!\!\!\!\!\!= {\cal Q}^{(2)}(\gamma_{10}=\frac{1}{2})=0\;,
	\\
	Q(EA) &=& Q(ACGE)
	\Big|_{\gamma_{21}=1} \!\!\!\!\!\!= {\cal Q}^{(2)}(\gamma_{10})\;,
	\label{equation:mad3conjecture1}
\end{eqnarray}
see \figref{fig:acge}.
To go beyond this initial observation
we fix $\gamma_{10}$, so that we can study the corresponding planar section of the parameter space in \figref{fig:MAD3-parameterSpace} which is depicted in~{\figref{fig:MAD3planar}}: here
the blue line belongs to the ABFE plane where
the capacity is known to equal to ${\cal Q}^{(2)}(\gamma_{10})$, while  the dashed purple line belongs instead to ACGE. From the monotonicity property of Corollary~\ref{corollary2}, we know that moving "up" or "right " in the $\gamma_{20}$, $\gamma_{21}$ depicted in \figref{fig:MAD3planar} can only decrease the quantum capacity value:
this in particular implies that all the points contained in the region R contained between
the ACGE and the ABFE plane
must have capacity larger than or equal to ${\cal Q}^{(2)}(\gamma_{10})$.
Notice next that from Eq.~\eqref{equation:mad3conjecture1} it follows that
at point $a$ of ACGE (i.e. for $\gamma_{21}=0$) the capacity is indeed
exactly equal to the lower bound ${\cal Q}^{(2)}(\gamma_{10})$.
This implies that at least for the points of R which can be reached from $a$ via
concatenation of "up" or "right" movement (purple triangle of the figure), must also have
capacity value given by~\eqref{equation:mad3conjecture}.
To extend this result to the remaining part of R, we focus on a new region in the parameter space, described by the relation:
\begin{equation}
	\label{equation:madgreen}
	\gamma_{21} = 1- k\gamma_{20}\quad k\in\left[1,2\right),
\end{equation}
which corresponds to the green line in \figref{fig:MAD3planar3} for a given value of $k$; the quantum capacity of MAD channels associated to this region is smaller than that of MAD's in the ACGE rectangle at fixed $\gamma_{10},\gamma_{20}$. Now, we try to find a channel $\Phi_{L}$ such that:
\begin{equation}
	\label{equation:connectingmad3}
	\Phi _{L} \circ\Phi _{\Gamma_1} = \Phi _{\Gamma_{2}},
\end{equation}
where $\Phi _{\Gamma_1},\Phi _{\Gamma_{2}}$ are MAD's corresponding to points in the purple and green regions depicted in \figref{fig:MAD3planar3} respectively. Using the inverse maps of MAD's \eqref{equation:definition:MADinverse} we observe that
the Choi matrix of $\Phi_{L}$ can be expressed as
\begin{widetext}
	\begin{equation}
		\label{equation:choiconnecting}
		\mathrm{Choi}\left(\Phi_{L}\right)=\tiny{\left(
			\begin{array}{ccccccccc}
				1                                                      & 0 & 0 & 0 & 1                                                      & 0 & 0                                                          & 0                                                 & \sqrt{\frac{2(k-1) (1-\omega_{21})}{k(1-\gamma_{21})}} \\
				0                                                      & 0 & 0 & 0 & 0                                                      & 0 & 0                                                          & 0                                                 & 0                                                      \\
				0                                                      & 0 & 0 & 0 & 0                                                      & 0 & 0                                                          & 0                                                 & 0                                                      \\
				0                                                      & 0 & 0 & 0 & 0                                                      & 0 & 0                                                          & 0                                                 & 0                                                      \\
				1                                                      & 0 & 0 & 0 & 1                                                      & 0 & 0                                                          & 0                                                 & \sqrt{\frac{2(k-1) (1-\omega_{21})}{k(1-\gamma_{21})}} \\
				0                                                      & 0 & 0 & 0 & 0                                                      & 0 & 0                                                          & 0                                                 & 0                                                      \\
				0                                                      & 0 & 0 & 0 & 0                                                      & 0 & \frac{2(1-\omega_{21})-k(1-\gamma_{21})}{k(1-\gamma_{21})} & 0                                                 & 0                                                      \\
				0                                                      & 0 & 0 & 0 & 0                                                      & 0 & 0                                                          & \frac{2 (\omega_{21}-\gamma_{21})}{1-\gamma_{21}} & 0                                                      \\
				\sqrt{\frac{2(k-1) (1-\omega_{21})}{k(1-\gamma_{21})}} & 0 & 0 & 0 & \sqrt{\frac{2(k-1) (1-\omega_{21})}{k(1-\gamma_{21})}} & 0 & 0                                                          & 0                                                 & \frac{2 (k-1) (1-\omega_{21})}{k(1-\gamma_{21})}       \\
			\end{array}
			\right)}
	\end{equation}
	The non-zero eigenvalues of this matrix are:
	\begin{equation}
		\label{equation:connectingeigen}
		\frac{2(1-\omega_{21})-k(1-\gamma_{21})}{k(1-\gamma_{21})},\quad\frac{2 (\omega_{21}-\gamma_{21})}{1-\gamma_{21}},\quad \frac{2k\left(2-\gamma_{21}-\omega_{21}\right)-2\left(1-\omega_{21}\right)}{k\left(1-\gamma_{21}\right)},
	\end{equation}
\end{widetext}
where the $\gamma_{ji}$'s are the transition probabilities of $\Phi_{\Gamma_{1}}$ while the $\omega_{ji}$'s are those of $\Phi_{\Gamma_{2}}$. Defining a sequence of decay probabilities:
\begin{equation}
	\label{equation:gammasequence}
	\gamma_{21}^{(n)} \coloneqq 1-\frac{1}{2^n},
\end{equation}
we find that the positivity of $\mathrm{Choi}\left(\Phi_{L}\right)$, under our conditions, reduces to:
\begin{equation}
	\label{equation:omegasequence0}
	\omega_{21}^{(n)} \leq 1-\frac{k}{2^{n+1}};
\end{equation}
note that $k$ can be as close to $1$ as possible, albeit always being bigger than $1$, so:
\begin{equation}
	\label{equation:omegasequence}
	\omega_{21}^{(n)} \leq 1-\frac{1}{2^{n+1}}.
\end{equation}
Using these sequences, we can do the following: we start from $\gamma_{21}^{(0)} =0$ and $\omega_{21}^{(0)}=1-k/2$, which correspond respectively to point $a,a'$ in \figref{fig:MAD3planar3}; there exists a LCPT map connecting these two points through \eqref{equation:connectingmad3}, therefore we get that $Q_a\geq Q_{a'}$, but at the same time $Q_{a'}\geq {\cal Q}^{(2)}(\gamma_{10})= Q_a$ as this is the quantum capacity of the region in the parameter space "above" $a'$ in \figref{fig:MAD3planar3}, so $Q_a = Q_{a'}$ for all $k\in [1,2)$, see \figref{fig:MAD3planar4}. Now we just need to iterate this process for $n>0$ and we find that the quantum capacity in the entire slice of \figref{fig:combinedmadplanar} is ${\cal Q}^{(2)}(\gamma_{10})$, proving \eqref{equation:mad3conjecture}.
\begin{figure*}
	\centering

	\begin{subfigure}[b]{0.45\textwidth}
		\centering
		\resizebox{1.0\textwidth}{!}{\begin{tikzpicture}[scale=6, >=stealth]
	\definecolor{mainblue}{RGB}{0, 114, 178}   
	\definecolor{mainpurple}{RGB}{128, 92, 156}
	\definecolor{shadecolor}{RGB}{173, 216, 230} 
	
	\coordinate (O) at (0,0);
	\coordinate (X1) at (1,0);
	\coordinate (Y1) at (0,1);
	\coordinate (A) at (0, 0.5);
	\coordinate (B) at (0.5, 0.5); 
	\coordinate (Xhalf) at (0.5, 0);
	\coordinate (Yhalf) at (0, 0.5);
	
	\draw[->, thick] (-0.05,0) -- (1.1,0) node[right] {\large $\gamma_{21}$};
	\draw[->, thick] (0,-0.05) -- (0,1.1) node[above] {\large $\gamma_{20}$};
	
	\draw (Xhalf) -- ++(0,-0.02) node[below] {$1/2$};
	\draw (X1) -- ++(0,-0.02) node[below] {$1$};
	\draw (Yhalf) -- ++(-0.02,0) node[left] {$1/2$};
	\draw (Y1) -- ++(-0.02,0) node[left] {$1$};
	
	\draw[mainblue, ultra thick] (Y1) -- (X1);
	
	\draw[black, thick, dotted] (A) -- (B);
	\draw[black, thick, dotted] (Xhalf) -- (B);
	
	\draw[mainpurple, ultra thick, dashed] (A) -- (X1);
	
	\node[circle, fill=mainpurple, inner sep=2pt, label={[black]below left:\large a}] at (A) {};
	\node[circle, fill=mainblue, inner sep=2pt, label={[black]below left:\large b}] at (B) {};
	
\end{tikzpicture}}
		\caption{The purple dashed line represent the ACGE rectangle in \figref{fig:MAD3-parameterSpace}, while the blue line represents the ABFE rectangle in the same figure.}
		\label{fig:MAD3planar}
	\end{subfigure}
	\hfill
	\begin{subfigure}[b]{0.45\textwidth}
		\centering
		\resizebox{1.0\textwidth}{!}{\begin{tikzpicture}[scale=6, >=stealth]
	\definecolor{mainblue}{RGB}{0, 114, 178}   
	\definecolor{mainpurple}{RGB}{128, 92, 156} 
	\definecolor{upperShade}{RGB}{128, 92, 156}  
	\definecolor{lowerShade}{RGB}{0, 60, 100}    
	
	\coordinate (O) at (0,0);
	\coordinate (X1) at (1,0);
	\coordinate (Y1) at (0,1);
	\coordinate (A) at (0, 0.5);
	\coordinate (B) at (0.5, 0.5); 
	\coordinate (Xhalf) at (0.5, 0);
	\coordinate (Yhalf) at (0, 0.5);
	
	\fill[upperShade, opacity=0.3] (A) -- (Y1) -- (B) -- cycle;
	
	\draw[->, thick] (-0.05,0) -- (1.1,0) node[right] {\large $\gamma_{21}$};
	\draw[->, thick] (0,-0.05) -- (0,1.1) node[above] {\large $\gamma_{20}$};
	
	\draw (Xhalf) -- ++(0,-0.02) node[below] {$1/2$};
	\draw (X1) -- ++(0,-0.02) node[below] {$1$};
	\draw (Yhalf) -- ++(-0.02,0) node[left] {$1/2$};
	\draw (Y1) -- ++(-0.02,0) node[left] {$1$};
	
	\draw[mainblue, ultra thick] (Y1) -- (X1);
	
	\draw[mainpurple, ultra thick, dashed] (A) -- (X1);
	
	\draw[black, thick, dotted] (A) -- (B);
	\draw[black, thick, dotted] (Xhalf) -- (B);
	
	\node[circle, fill=mainpurple, inner sep=2pt, label={[black]below left:\large a}] at (A) {};
	\node[circle, fill=mainblue, inner sep=2pt, label={[black]below left:\large b}] at (B) {};
	
	\node[mainpurple] at (0.17, 0.65) {${\cal Q}^{(2)}(\gamma_{10})$};
	\node[mainblue, anchor=south west] at (0.4, 0.6) {${\cal Q}^{(2)}(\gamma_{10})$};
	
	\draw[<-, thick, mainpurple] (A) ++(-0.02, 0.02) -- ++(-0.17, 0.15) node[above, mainpurple] {${\cal Q}^{(2)}(\gamma_{10})$};
	
	\node[black] at (0.95, 0.45) {\large $Q_a = Q_b$};
	
\end{tikzpicture}}
		\caption{Using the monotonicity property of Corollary~\ref{corollary2}, we can infer the quantum capacity in the upper purple triangle.}
		\label{fig:MAD3planar2}
	\end{subfigure}

	\vspace{2em}

	\begin{subfigure}[b]{0.45\textwidth}
		\centering
		\resizebox{1.0\textwidth}{!}{\begin{tikzpicture}[scale=6, >=stealth]
	\definecolor{mainblue}{RGB}{0, 114, 178}   
	\definecolor{mainpurple}{RGB}{128, 92, 156} 
	\definecolor{maingreen}{RGB}{34, 139, 34}   
	\definecolor{axisGray}{RGB}{80, 80, 80}
	
	\coordinate (O) at (0,0);
	\coordinate (X1) at (1,0);
	\coordinate (Y1) at (0,1);
	\coordinate (A) at (0, 0.5);
	\coordinate (Aprime) at (0.35, 0.45);
	\coordinate (B) at (0.5, 0.5); 
	\coordinate (Xhalf) at (0.5, 0);
	\coordinate (Yhalf) at (0, 0.5);
	
	\draw[thick, ->, axisGray] (-0.1,0) -- (1.2,0) node[right, black] {\large $\gamma_{21}$};
	\draw[thick, ->, axisGray] (0,-0.1) -- (0,1.1) node[above, black] {\large $\gamma_{20}$};
	
	\draw (Xhalf) -- ++(0,-0.02) node[below] {$1/2$};
	\draw (X1) -- ++(0,-0.02) node[below] {$1$};
	\draw (Yhalf) -- ++(-0.02,0) node[above left] {$1/2$};
	\draw (Y1) -- ++(-0.02,0) node[left] {$1$};
	
	\draw[mainblue, ultra thick] (0,1) -- (1,0);
	
	\draw[mainpurple, ultra thick, dashed] (0, 0.5) -- (1, 0) 
	node[pos=0.3, below, xshift=-5pt] {};
	
	\draw[maingreen, ultra thick, dashed] (0, 0.7) -- (1, 0) 
	node[pos=0.4, above, xshift=5pt] {};
	
	\draw[black, thick, dotted] (0, 0.5) -- (0.5, 0.5);
	\draw[black, thick, dotted] (0.5, 0) -- (0.5, 0.8);
	
	\node[circle, draw=mainpurple, fill=white, inner sep=1.5pt, thick] at (0.5, 0.5) {};
	
	\draw[->, red, ultra thick] (A) -- (Aprime) 
	node[midway, above left] {\large $\phi_L$};
	
	\node[anchor=west] at (0.6, 0.7) {
		\large ${\color{red}{\phi_L}} \circ {\color{mainpurple}\phi_{\Gamma_a}}= {\color{maingreen}\phi_{\Gamma_{a'}}}$
	};
	
	
	\node[circle, fill=mainpurple, inner sep=2pt, label={[mainpurple]below left:\large a}] at (A) {};
	\node[circle, fill=maingreen, inner sep=2pt, label={[maingreen]below:\large a'}] at (Aprime) {};
	
\end{tikzpicture}}
		\caption{The connecting channel in \eqref{equation:connectingmad3} can be used to connect the two points $a,a'$, given by \eqref{equation:gammasequence} and \eqref{equation:omegasequence0} respectively for $n=0$ and for a value of $k\in[1,2)$.}
		\label{fig:MAD3planar3}
	\end{subfigure}
	\hfill
	\begin{subfigure}[b]{0.45\textwidth}
		\centering
		\resizebox{1.0\textwidth}{!}{\begin{tikzpicture}[scale=6, >=stealth]
	\definecolor{mainblue}{RGB}{0, 114, 178}    
	\definecolor{mainpurple}{RGB}{128, 92, 156}  
	\definecolor{upperShade}{RGB}{128, 92, 156}  
	\definecolor{lowerShade}{RGB}{0, 60, 100}    
	\definecolor{axisGray}{RGB}{80, 80, 80}
	
	\coordinate (O) at (0,0);
	\coordinate (X1) at (1,0);
	\coordinate (Y1) at (0,1);
	\coordinate (A) at (0, 0.5);
	\coordinate (B) at (0.5, 0.5);
	\coordinate (Xhalf) at (0.5, 0);
	\coordinate (Yhalf) at (0, 0.5);
	
	\fill[upperShade, opacity=0.3] (A) -- (Y1) -- (B) -- cycle;
	
	\fill[lowerShade, opacity=0.6] (A) -- (B) -- (0.5, 0.25) -- cycle;
	
	\draw[thick, ->, axisGray] (-0.1,0) -- (1.2,0) node[right, black] {\large $\gamma_{21}$};
	\draw[thick, ->, axisGray] (0,-0.1) -- (0,1.1) node[above, black] {\large $\gamma_{20}$};
	
	\draw (Xhalf) -- ++(0,-0.02) node[below] {\large $1/2$};
	\draw (X1) -- ++(0,-0.02) node[below] {\large $1$};
	\draw (Yhalf) -- ++(-0.02,0) node[left] {\large $1/2$};
	\draw (Y1) -- ++(-0.02,0) node[left] {\large $1$};
	
	\draw[mainblue, ultra thick] (Y1) -- (X1);
	
	\draw[mainpurple, ultra thick, dashed] (A) -- (X1);
	
	\draw[black, thick, dotted] (A) -- (B);
	\draw[black, thick, dotted] (Xhalf) -- (Xhalf |- 0, 0.9);
	
	\node[circle, draw=mainpurple, fill=white, inner sep=1.5pt, thick] at (B) {};
	
	\node[mainpurple] at (0.17, 0.65) {${\cal Q}^{(2)}(\gamma_{10})$};
	\node[white] at (0.32, 0.43) {\footnotesize ${\cal Q}^{(2)}(\gamma_{10})$};
	
\end{tikzpicture}}
		\caption{Varying the value of $k$, the connecting map in \figref{fig:MAD3planar3} spans all the points in the lower blue triangle, allowing us to infer the value of the quantum capacity in that region.}
		\label{fig:MAD3planar4}
	\end{subfigure}

	\caption{Slice of the parameter space in \figref{fig:MAD3-parameterSpace} at fixed $\gamma_{10}$. These figures illustrate the first step in the algorithm used to find the quantum capacity in the entire ABCEFG cheese-wedge of \figref{fig:MAD3-parameterSpace}.}
	\label{fig:combinedmadplanar}
\end{figure*}






\section{Conclusions}
\label{section:conclusions}

We presented a systematic characterazation of MAD's, which are able to model energy decay processes in finite-dimensional systems. We found general results that can be applied at any dimension (Section \ref{section:setting}) and found a working recipe for the computation of the quantum capacity (Section \ref{section:procedure}) in non-degradle regions (note that these channels are not subadditive, so the single-letter formula for the quantum capacity can not be used in non-degradable regions). Theorem \ref{theorem1} has direct applications in real-world scenarios where the decay rates present a dependency on the distance of the transmission: it tells us how far we can communicate before having to build a \textit{repeater}. As a sandbox for our results, we computed the quantum capacity of $4$-dimensional MAD's in some regions of their parameter space(Section \ref{section:example}), making heavy use of the Theorem \ref{Teorema2}. Lastly, we formulate a conjecture on the optimal encoding of MAD's when only some of the antidegradability conditions are satisfied (see later, \ref{subsection:conjecture}); trusting the conjecture, we were also able to compute the quantum capacity on new regions of the parameter space of $3$-dimensional MAD's not yet explored (Subsection \ref{subsubsection:mad3}).

We are grateful for the insightful discussions with L. Lami, V. Vetrano, S. Chessa (who also provided the first version of the code used for the computations of the quantum capacity) and F. A. Mele (who introduced us to SDP's and provided helpful support in their application).

Finally, we acknowledge the financial support by MUR (Ministero dell'Università e della Ricerca) through the PNRR MUR project PE0000023-NQSTI.

\bibliography{biblio.bib}

\newpage

\appendix
\renewcommand{\thesection}{A.\arabic{section}}
\section*{Appendix}
\section{About composition rules of MAD channels}
\label{appendix:comprules}
This section contains technical material associated to Sec.~\ref{section:setting} of the main text.

\subsection{Convex combination of MAD channels}

Here we show that the convex combinations of MAD channels correspond to a MAD channel up to a dephasing transformation as shown in Eq.~\eqref{convexityconnection}.
The transition elements of the channel $\Phi_{\Gamma(p)}$ are
\begin{eqnarray}
	\gamma_{ji}(p) :=p \gamma_{ji} + (1-p) \gamma'_{ji}\,.
\end{eqnarray}
It turns out that
these coefficients determine the evolution of the population terms induced by the convex combination of MAD channels, indeed from
Eq.~\eqref{equation:definition:madchannelkraus1} it follows
\begin{equation}
	\label{equation:definition:madchannelkraus1conv1}
	\left(p \Phi_{\Gamma} + (1-p )\Phi_{\Gamma'}  \right)(|j\rangle\langle j|)
	=\sum_{i'=0}^{j} \gamma_{ji}(p) |i'\rangle\langle i'| \;.	\end{equation}
which perfectly mimics the action of
$\Phi_{\Gamma(p)}$.
On the contrary for the off-diagonal entries we get
\begin{eqnarray}
	\label{equation:definition:madchannelkraus1conv}
	&&\left(p \Phi_{\Gamma} + (1-p )\Phi_{\Gamma'}  \right)(|j\rangle\langle i|)
	\\\nonumber
	&& \qquad = \left(p \sqrt{\gamma _{jj}\gamma_{ii}}+ (1-p )
	\sqrt{\gamma'_{jj}\gamma'_{ii}}\right) |j\rangle\langle i| \nonumber \\
	&& \qquad \neq \nonumber 	\sqrt{\gamma_{jj}(p)\gamma_{ii}(p)} |j\rangle\langle i| =\Phi_{\Gamma(p)}	(|j\rangle\langle i|)\;.\end{eqnarray}
Observe next that   the parameters
\begin{equation}
	\eta_{ji} =\eta_{ij}:=\frac{p \sqrt{\gamma _{jj}\gamma_{ii}}+ (1-p )
	\sqrt{\gamma'_{jj}\gamma'_{ii}}}{
	\sqrt{\gamma_{jj}(p)\gamma_{ii}(p)}}\;,\end{equation}
belong to the interval $[0,1]$, with diagonal entries that are always equal to 1. Accordingly we can identify the dephasing channel in Eq.~\eqref{convexityconnection} with the map
\begin{eqnarray}
	\Delta(|j\rangle\langle i|) =
	\eta_{ji} |j\rangle\langle i| \;.
\end{eqnarray}

\subsection{Proof of Eq.~ (\ref{equation:MADcomposition})}
Let's first notice that the product  of two lower-triangular matrixes is also lower a lower-triangular.
Furthermore since for fixed $j$,  the matrix elements
$\{\gamma ^\prime _{ji}\}_{i=0,\cdots, j}$ of
$\Gamma'$ and  $\{\gamma ^{\prime\prime} _{ji}\}_{i=0,\cdots, j}$  of $\Gamma''$,  are conditional probabilities it follows that   also the matrix elements   $\{\gamma _{ji}=\sum_{\ell =i} ^{j}\gamma ^\prime _{j\ell} \gamma^{\prime\prime} _{\ell i}\}_{i=0,\cdots, j}$ of $\Gamma$ have the same property implying that they fulfil the transition matrix constraints (\ref{equation:gammajiproperties}). Next goal is to show that indeed $\Phi _{\Gamma }=\Phi _{\Gamma ^{\prime\prime}}\circ\Phi _{\Gamma ^\prime}$.
This can be proved by invoking   Eq.~(\ref{equation:definition:madchannelkraus1}) and observing that
\begin{eqnarray}
	\label{equation:definition:madchannelkraus2}
	\Phi _{\Gamma ^{\prime\prime}}\circ\Phi _{\Gamma ^\prime} (|j\rangle\langle i|) 	&=&\Phi _{\Gamma ^{\prime\prime}}(\Phi _{\Gamma ^\prime} (|j\rangle\langle i|)) \\
	&=&\nonumber
	\sqrt{\gamma^{\prime\prime} _{jj}\gamma^{\prime\prime}_{ii}\gamma^{\prime} _{jj}\gamma^{\prime}_{ii}} \;\Phi _{\Gamma ^{\prime\prime}}(|j\rangle\langle i|) 		\\ &=&\sqrt{\gamma_{jj}\gamma_{ii}}\; |j\rangle\langle i| =
	\Phi _{\Gamma }(|j\rangle\langle i|)  \nonumber  \;,
\end{eqnarray}
for $j\neq i$ and
\begin{eqnarray}\nonumber
	&&\Phi _{\Gamma ^{\prime\prime}}\circ\Phi _{\Gamma ^{\prime}} (|j\rangle\langle j|) 	=
	\sum_{i'=0}^{j}\gamma^{\prime}_{ji'}  \; \Phi _{\Gamma ^{\prime\prime}} (|i'\rangle\langle i' |) 	\\  &&\quad = \label{equation:definition:madchannelkraus3}
	\sum_{i'=0}^{j}\sum_{i''=0}^{i'}\gamma^{\prime\prime}_{i' i''}\gamma^{\prime}_{ji'} \; |i''\rangle\langle i''|
	\\  	&&\quad= \nonumber
	\sum_{i'=0}^{j}\sum_{i''=0}^{i'}\gamma_{j i''} \; |i''\rangle\langle i''|
	=\Phi _{\Gamma} (|j\rangle\langle j|) \;.
\end{eqnarray}
Notice finally that the fully damping channel $\Phi_{\Gamma_{\emptyset}}$ of~\eqref{equation:definition:FULLY}
represents a zero element for the group of MAD channels, thanks to the fact that
\begin{equation}
	\Gamma_{\emptyset} \Gamma =
	\Gamma \Gamma_{\emptyset} =  \Gamma_{\emptyset}\,,  \quad \forall \Gamma \in{\cal M}_d\;,
\end{equation}
which implies
\begin{equation}
	\Phi_{\Gamma_{\emptyset}}\circ  \Phi_{\Gamma} =
	\Phi_{\Gamma} \circ \Phi_{\Gamma_{\emptyset}}=  \Phi_{\Gamma_{\emptyset}}\,.
\end{equation}

\subsection{Single-decays decomposition}
\label{appendix:isolatedecay}
\begin{figure*}
	\centering
	\scalebox{1}{
		\begin{tikzpicture}[scale=1.5]
	\draw[black, thick] (2.5,0) -- (0,0) node [left] {\small{$\ket{0}$}};
	\draw[black, thick] (2.5,0.5) -- (0,0.5) node [left] {\small{$\ket{1}$}};
	\draw[black, thick] (2.5,1) -- (0,1) node[left] {\small{$\ket{2}$}};
	\draw[black, thick] (2.5,1.5) -- (0,1.5) node [left] {\small{$\ket{3}$}};
	\draw[black, thick, ->] (1,1.5) -- (1,1) node [above left] {\footnotesize{$\gamma_{32}$}};
	\draw[black, thick, ->] (1.5,1.5) -- (1.5,0.5) node [above left] {\footnotesize{$\gamma_{31}$}};
	\draw[black, thick, ->] (2,1.5) -- (2,0) node [above left] {\footnotesize{$\gamma_{30}$}};
	
	\node (sim) at (2.8,0.75) {\small{$\sim$}};
	
	\draw[black, thick] (5,0) -- (3.5,0) node [left] {\small{$\ket{0}$}};
	\draw[black, thick] (5,0.5) -- (3.5,0.5) node [left] {\small{$\ket{1}$}};
	\draw[black, thick] (5,1) -- (3.5,1) node[left] {\small{$\ket{2}$}};
	\draw[black, thick] (5,1.5) -- (3.5,1.5) node [left] {\small{$\ket{3}$}};
	
	\draw[black, thick, ->] (4,1.5) -- (4,1) node [above left] {\footnotesize{$\gamma _{32}$}};
	\draw[black, thick, ->] (4.5,1.5) -- (4.5,0) node [above left] {\footnotesize{$\gamma _{30}$}};
	
	\node (then) at (5.3,0.75) {\small{$\rightarrow$}};
	
	\draw[black, thick] (7.5,0) -- (6,0) node [left] {\small{$\ket{0}$}};
	\draw[black, thick] (7.5,0.5) -- (6,0.5) node [left] {\small{$\ket{1}$}};
	\draw[black, thick] (7.5,1) -- (6,1) node[left] {\small{$\ket{2}$}};
	\draw[black, thick] (7.5,1.5) -- (6,1.5) node [left] {\small{$\ket{3}$}};
	
	\draw[black, thick, ->] (6.75,1.5) -- (6.75,0.5) node [above left] {\footnotesize{$\gamma_{31}^{\prime}$}};
	
\end{tikzpicture}
	}
	\caption{Example of the decomposition in \eqref{equation:composition:leftisolation2}
		, having fixed $k=3,n=1$ for $4$-dimensional MAD channels, read from left to right in "chronological" order. In this case, \eqref{equation:composition:leftisolation2}
		implies that the $4$-dimensional MAD channel representing decays from level $\ket{3}$ onto lower levels can be decomposed by isolating a single decay, which is performed after a transition encompassing all the other decays and with a modified transition probability, defined in \eqref{equation:composition:leftisolation}.}
	\label{fig:decomposition2}
\end{figure*}
Given $\Gamma_k$  as in \eqref{equation:gammadeco1} and $	\Xi _k ^{(n)}$ as in
\eqref{equation:definition:gammakdecotext}, we observe that, setting
\begin{equation}
	\label{equation:composition:leftisolation}
	\xi _{kn} \coloneqq  \frac{\gamma _{kn}}{1-\sum _{\substack{i=0\\i\neq n}} ^{k-1} \gamma _{ki}}\;,
\end{equation}
they following identity holds
\begin{eqnarray} \label{equation:composition:leftisolation1}
	\Gamma _k &= T_k ^{(n)} 	\Xi _k ^{(n)},
\end{eqnarray}
where $T_k ^{(n)}$ is the transition matrix obtained by removing the decay from $k$ to $n$ from
$\Gamma _k$, i.e.
\begin{eqnarray}
	T _k ^{(n)} &\coloneqq&\Gamma _k -\gamma _{kn}\op{k}{n}+\gamma_{kn}\op{k}{k}
	\label{equation:gammakdeco1}\\ \nonumber
	&=&  \mathds{1}+ \sum _{\substack{i=0\\i\neq n}} ^{k-1} \gamma _{ki} \op{k}{i} -\sum _{\substack{i=0\\i\neq n}} ^{k-1} \gamma _{ki} \op{k}{k}\;.
\end{eqnarray}
At the level of MAD channels Eq.~(\ref{equation:composition:leftisolation1}) translates into
\begin{eqnarray} \label{equation:composition:leftisolation2}
	\Phi_{\Gamma _k} = \Phi_{\Xi _k ^{(n)}}\circ
	\Phi_{T_k ^{(n)}} \;,
\end{eqnarray}
whose physical interpretation is clear: given all possible decays from the level $\ket{k}$, we can think of this process chronologically and choose the decay onto $\ket{n}$ to be performed at the end, with modified transition probabilities, as in \eqref{equation:composition:leftisolation}; a schematic graph of this decomposition is reported in \figref{fig:decomposition2}.
The decomposition~\eqref{equation:composition:leftisolation1} can be expanded even further in order to create a composition of single-decay matrices. This can be achieved by a slight redefinition of the matrix in \eqref{equation:gammakdeco1}:
\begin{equation}
	\label{equation:definition:gammakdeco}
	H ^{(n)} _k \coloneqq \mathds{1}+ \sum _{i=n} ^{k-1} \gamma _{ki} \op{k}{i} -\sum _{i=n} ^{k-1} \gamma _{ki} \op{k}{k}\;,
\end{equation}
which represents a MAD channel where only the $\ket{k}$ level can decay, with the same amplitudes as the original matrix $\Gamma_k$ and having the transitions onto the lowest $n$ levels forbidden. One can verify that:
\begin{equation}
	\label{equation:singleleveldeco1}
	H^{(0)}_{k} = \Gamma _k\;,
\end{equation}
which matches the descriptive definition given above.
Furthermore taking $\Xi _k ^{(n)}$ with transition probability
\begin{eqnarray}
	\xi_{kn}&\coloneqq& \frac{\gamma _{kn}}{1-\sum _{i=n+1} ^{k-1}\gamma_{ki}}
	\label{equation:definition:modifiedamplitudes}\\
	\nonumber &=& \frac{\gamma_{kn}}{\gamma_{kn}+(\gamma_{kk}+\sum_{i=0}^{n-1}\gamma_{ki})} \;,
\end{eqnarray}
we can write
\begin{equation}
	\label{equation:definition:decogammakiter}
	H_k ^{(n)} = H_k ^{(n+1)} \Xi _k ^{(n)},
\end{equation}
which for $n=k-1$ reduces to
\begin{equation}
	\label{equation:singleleveldeco2}
	H _k ^{(k-1)}=\Xi _k ^{(k-1)}.
\end{equation}
Thus, invoking (\ref{equation:singleleveldeco1}) we  get
\begin{eqnarray}
	\label{equation:GammakADCdeco}
	\Gamma _k &=& H ^{(0)}=H_k ^{(1)}\Xi _k ^{(0)}=H_k ^{(2)}\Xi _k ^{(1)}\Xi _k ^{(0)}\nonumber\\
	&=&\cdots = \Xi _k ^{(k-1)} \cdots \Xi _k ^{(0)}.
\end{eqnarray}
This means that the MAD channel that allows for decays from a single level can be decomposed into a series of single-decay MAD's with appropriately modified transition amplitudes.
Applying this to Eq.~(\ref{equation:GammaSingleLevelDeco}) we finally arrive to (\ref{equation:GammaADCdeco}) and hence (\ref{equation:MADADCdeco}).

\section{Degradable channels decompositions} \label{appendix:degradabledecomposition}

In this section we provide proof of Eq.~\eqref{equation:qcapacitydeg2}, i.e. that
if the composite LCPTP channel $\Psi=\Phi\circ \Phi'$ is degradable and $\Phi'$ is invertible, then
also the LCPTP channel $\Phi$ is also degradable.

Indicating with $\mathcal{H}_A$ and $\mathcal{H}_{A'}$ the input and output Hilbert spaces  of  $\Phi'$, and with $\mathcal{H}_{A'}$ and $\mathcal{H}_{B}$ the input and output Hilbert spaces  of  $\Phi$, consider the Stinespring dilations \cite{StinespringDilation} of these channels:
\begin{eqnarray}
	\Phi' \leftrightarrow V'&:& \mathcal{H}_A\mapsto \mathcal{H}_{A'}\otimes \mathcal{H}_{E'},\nonumber\\
	\Phi \leftrightarrow V&:&\mathcal{H}_{A'}\mapsto \mathcal{H}_{B}\otimes \mathcal{H}_{E},\nonumber
\end{eqnarray}
with $\mathcal{H}_{E'}$ and $\mathcal{H}_{E}$ independent environmental Hilbert spaces, and
$V'$ and $V$ corresponding isomorphisms.
By construction we can hence identify the isomorphism
\begin{equation}
	\left(V\otimes\mathrm{Id}_{E'}\right)V': \mathcal{H}_A\mapsto \mathcal{H}_{B}\otimes \mathcal{H}_{E'}\otimes \mathcal{H}_{E},
\end{equation}
as a valid Stinespring dilation of $\Psi$, so that, for all inputs state $\rho$ of $\mathcal{H}_A$ we can write
\begin{equation}
	\label{equation:stinespring}
	{\Psi} (\rho) = \tr_{E'E}\left[\left(V\otimes\mathrm{Id}_{E'}\right)V' \rho V'^\dagger\left(V^\dagger\otimes\mathrm{Id}_{E'}\right)\right],
\end{equation}
and
\begin{equation}
	\label{equation:complementary_stinespring}
	\tilde{\Psi} (\rho) = \tr_{B}\left[\left(V\otimes\mathrm{Id}_{E'}\right)V' \rho V'^\dagger\left(V^\dagger\otimes\mathrm{Id}_{E'}\right)\right].
\end{equation}
By direct computation, it is also easy to show that:
\begin{eqnarray}
	\label{equation:complementary_traces}
	\tr_{E}\tilde{\Psi} (\rho) &=&  \tilde{\Phi}' (\rho),\nonumber\\
	\tr_{E'}\tilde{\Psi} (\rho) &=& \tilde{\Phi} \left(\Phi'(\rho)\right),
\end{eqnarray}
where $ \tilde{\Phi}$ and $\tilde{\Phi}'$ are respectively the complementary channels of $\Phi$ and $\Phi'$.
If $\Psi$ is degradable there must exist an LCPTP map $\Lambda$ form $\mathcal{H}_B$ to
$\mathcal{H}_{E'}\otimes \mathcal{H}_{E}$ such that
\begin{eqnarray}
	\label{equation:degpsi2}
	\tilde{\Psi} (\rho)= \Lambda\circ \Psi(\rho)&=&\Lambda\circ \Phi\circ \Phi' (\rho)\\
	&=&(\Lambda\circ \Phi)( \Phi' (\rho)) \;. \nonumber
\end{eqnarray}
Under  partial trace w.r.t. $E'$ this leads to
\begin{eqnarray}
	\label{equation:complementary_trace1}
	\tilde{\Phi} \left(\Phi'(\rho)\right)&=& \tr _{E'} [(\Lambda\circ \Phi)(\Phi' (\rho))] \\
	&=&(\Lambda'\circ \Phi)(\Phi' (\rho))\;, \nonumber
\end{eqnarray}
where we defined the LCPTP map
\begin{equation} \Lambda' \coloneqq\tr _{E'}\circ\Lambda \;.
\end{equation}
Now, if $\Phi'$ is invertible, we can apply such identity to states of form
$\tau=(\Phi' )^{-1}(\rho)$ obtaining
\begin{equation}
	\label{equation:degpsi2_1}
	\tilde{\Phi}'(\rho) = \Lambda'\circ \Phi'(\rho),
\end{equation}
which being valid for all $\rho$ proves the
degradability of $\Phi'$.
\section{Quantum capacity of DS channels with entanglement breaking components}\label{APPDS}
In this section we prove Eq.~\eqref{equation:dscapacity1}, showing that the same identity holds for all
DS channel~\eqref{generalDS} whose component
$\Phi_{YY'}$ is an Entanglement Breaking (EB) channel~\cite{holevo_book}.
Let's first recall that EB channels are measure and re-prepare transformations and observe that
${\Delta}_{{\mathcal{A}'}}$
of Eq.~\ref{DeltaAprime} belongs to such class of transformations once
we identify the measurement  with a projection  w.r.t. to the computational basis of the space  ${\mathcal{A}'}$.
A notable property of EB channels is that when employed in conjunction of  others channel $\Phi$, they cannot increase the quantum
capacity of the communication line~\cite{Barnum2000,Bennett2002,mad3}, i.e.
\begin{equation}
	\label{equation:qcapacitydegEBassisted}
	Q\left(\Phi \otimes \Phi^{\rm EB}\right) = Q\left(\Phi \right)\;,
\end{equation}
which in particular implies  that
\begin{equation}
	\label{equation:qcapacitydegMADn}
	I_c \left(\rho,\Phi \otimes \Phi^{\rm EB}\right) \leq  Q\left(\Phi \right)\;,
\end{equation}
for all input states $\rho$ of the composite channel.
\begin{lemma}\label{lemmaAPP}
	Let $\Phi^{\rm DS}=\Phi_{XX'}\oplus \Phi_{YY'}$ be a DS channel from
	${\mathcal{H}}_Z=
			{\mathcal{H}}_X \oplus {\mathcal{H}}_Y$ to
	${\mathcal{H}}_{Z'}=
		{\mathcal{H}}_{X'} \oplus {\mathcal{H}}_{Y'}$, with
	$\Phi_{YY'}$ being EB.
	Then the quantum capacity of $\Phi^{\rm DS}$ coincides with
	the quantum capacity of $\Phi_{XX'}$ alone, i.e.
	\begin{equation}
		\label{equation:dscapacity1gen}
		Q(\Phi^{\rm DS})=Q(\Phi_{XX'})\;.
	\end{equation}
\end{lemma}
\begin{proof}
	The capacity of  $\Phi^{\rm DS}$ is not smaller than the capacity
	of $\Phi_{XX'}$, since this communication rate can be achieved by
	only coding the input messages on ${\mathcal{H}}_X$.
	\begin{equation}
		\label{inequation:dscapacity1gen}
		Q(\Phi^{\rm DS}) \geq Q(\Phi_{XX'})\;.
	\end{equation}
	To prove the thesis we hence need to show that also the opposite
	inequality holds.
	To show this let us  observe that  for all $n$ integers ${\mathcal{H}}^{\otimes n}_Z$
	is given by the direct sum of all subspaces obtained by taking the tensor
	product of $n-k$ copies of ${\mathcal{H}}_X$ and
	$k$ copies of ${\mathcal{H}}_Y$ for all $k=0, \cdots, n$, i.e.
	\begin{eqnarray}
		{\mathcal{H}}^{\otimes n}_Z&=&{\mathcal{H}}_{Z_1}\otimes
		{\mathcal{H}}_{Z_2}\otimes \cdots \otimes {\mathcal{H}}_{Z_n}\nonumber\\
		&=&({\mathcal{H}}_{X_1} \oplus\nonumber
		{\mathcal{H}}_{Y_1})\otimes  \cdots \otimes
		({\mathcal{H}}_{X_n} \oplus
		{\mathcal{H}}_{Y_n})\\
		&=&\bigoplus_{k=0}^n \left(\bigoplus_{\ell=1}^{\binom{n}{k}} {\mathcal{H}}^{(n,k;\ell)}_{XY}\right)\;,\end{eqnarray}
	where for fixed $n$ and $k$,  ${\mathcal{H}}^{(n,k;1)}_{XY}$,  ${\mathcal{H}}^{(n,k;2)}_{XY}$, $\cdots$,  ${\mathcal{H}}^{(n,k;)}_{XY}$ are the $\binom{n}{k}$ subspaces of
	${\mathcal{H}}^{\otimes n}_{Z}$ that includes
	that involve $n-k$ copies of ${\mathcal{H}}_{X}$ and $k$ copies of
	${\mathcal{H}}_{Y}$, e.g.
	\begin{eqnarray*}
		{\mathcal{H}}^{(2,0;1)}_{XY}&=&{\mathcal{H}}_{X_1}\otimes
		{\mathcal{H}}_{X_2} \;, \\
		{\mathcal{H}}^{(2,1;1)}_{XY}&=&{\mathcal{H}}_{X_1}\otimes
		{\mathcal{H}}_{Y_2}\;,\\
		{\mathcal{H}}^{(2,1;2)}_{XY}&=& {\mathcal{H}}_{X_2}\otimes
		{\mathcal{H}}_{Y_1} \;,\\ {\mathcal{H}}^{(2,2;1)}_{XY}&=&{\mathcal{H}}_{Y_1}\otimes
		{\mathcal{H}}_{Y_2}\;.
	\end{eqnarray*}
	The same expansion can be adopted for the output space ${\mathcal{H}}^{\otimes n}_{Z'}$, i.e.
	\begin{eqnarray}
		{\mathcal{H}}^{\otimes n}_{Z'}&=&\bigoplus_{k=0}^n \left(\bigoplus_{\ell=1}^{\binom{n}{k}} {\mathcal{H}}^{(n,k;\ell)}_{X'Y'}\right)\;,\end{eqnarray}
	where for  fixed $n$, $k$, $\ell$,
	${\mathcal{H}}^{(n,k;\ell)}_{X'Y'}$ is formed by the  output  spaces associated with those
	that define  ${\mathcal{H}}^{(n,k;\ell)}_{XY}$, e.g.
	\begin{eqnarray*}
		{\mathcal{H}}^{(2,0;1)}_{X'Y'}&=&{\mathcal{H}}_{X'_1}\otimes
		{\mathcal{H}}_{X'_2} \;, \\
		{\mathcal{H}}^{(2,1;1)}_{X'Y'}&=&{\mathcal{H}}_{X'_1}\otimes
		{\mathcal{H}}_{Y'_2}\;,\\
		{\mathcal{H}}^{(2,1;2)}_{X'Y'}&=& {\mathcal{H}}_{X'_2}\otimes
		{\mathcal{H}}_{Y'_1} \;,\\ {\mathcal{H}}^{(2,2;1)}_{X'Y'}&=&{\mathcal{H}}_{Y'_1}\otimes
		{\mathcal{H}}_{Y'_2}\;.
	\end{eqnarray*}
	With this notation we can now express
	$(\Phi^{\rm DS})^{\otimes n}$ as a direct sum of channels $\Phi^{(n,k;\ell)}$
	that maps ${\mathcal{H}}^{(n,k;\ell)}_{XY}$  into the corresponding output space ${\mathcal{H}}^{(n,k;\ell)}_{X'Y'}$ via a tensor product of transformations
	$\Phi_{XX'}$ and   $\Phi_{YY'}$. Specifically
	\begin{eqnarray}\label{generalDSn}
		(\Phi^{\rm DS})^{\otimes n} &=&
		\bigoplus_{k=0}^n \bigoplus_{\ell=1}^{\binom{n}{k}} \Phi^{(n,k;\ell)}\;,
	\end{eqnarray}
	with
	\begin{eqnarray}
		\Phi^{(n,k;\ell)}:=\Phi_{XX'}^{(n,k;\ell)}\otimes \Phi_{YY'}^{(n,k;\ell)}\;,\end{eqnarray}
	and
	\begin{equation}
		\Phi_{XX'}^{(n,k;\ell)}:=	\bigotimes_{j} \Phi_{X_jX_j'} \;,
		\quad
		\Phi_{YY'}^{(n,k;\ell)}:= \bigotimes_{i}\Phi_{Y_i Y_i'}\;,
	\end{equation}
	where in the last expressions the indexes $j$ and $i$ run, respectively, on all the $n-k$ copies of
	${\mathcal{H}}_{X}$ and the $k$ copies of ${\mathcal{H}}_{Y}$,  contained in ${\mathcal{H}}^{(n,k;\ell)}_{XY}$.
	We then recall that  the coherent information of  DS channels can be expressed as the maximum of the coherent information of its addenda~\cite{direct_sum_channels}. In our case this translates into the following inequalities
	\begin{eqnarray}\nonumber
		&&I_c \left(\rho^{(n)},(\Phi^{\rm DS})^{\otimes n} \right) =
		\max_{k,\ell}\left\{  I_c \left(\rho^{(n)},\Phi^{(n,k;\ell)} \right)\right\}\\ &&\quad = \max_{k,\ell}\left\{  I_c \left(\rho^{(n)},\Phi_{XX'}^{(n,k;\ell)}\otimes \Phi_{YY'}^{(n,k;\ell)}\right) \right\}\nonumber \\
		&&\quad  \leq\max_{k,\ell}\left\{  Q\left(\Phi_{XX'}^{(n,k;\ell)} \right)\right\}\label{equation:qcapacitydegMADn1}
	\end{eqnarray}
	where in the last step we used~\eqref{equation:qcapacitydegMADn} and  the fact that $Phi_{YY'}^{(n,k;\ell)}$ is EB. Observe next that, by definition  $\Phi_{XX'}^{(n,k;\ell)}$ is the tensor product of $n-k$ copies of the channel $\Phi_{XX'}$, so that
	\begin{eqnarray}\nonumber
		Q\left(\Phi_{XX'}^{(n,k;\ell)} \right)=
		Q\left(\Phi_{XX'}^{\otimes (n-k)} \right)&=&
		(n-k)Q\left(\Phi_{XX'} \right)\nonumber \\
		&\leq&n Q\left(\Phi_{XX'} \right) \;, \nonumber
	\end{eqnarray}
	which inserted in \eqref{equation:qcapacitydegMADn1} leads to
	\begin{eqnarray}\nonumber
		&&\frac{I_c \left(\rho^{(n)},(\Phi^{\rm DS})^{\otimes n} \right)}{n}   \leq
		Q\left(\Phi_{XX'} \right) \label{equation:qcapacitydegMADn2}\;.
	\end{eqnarray}
	Taking the supremum  w.r.t. to all possible inputs $\rho^{(n)}$, and then limit $n\rightarrow \infty$, the left-hand-side term converges to
	$Q\left(\Phi^{\rm DS}\right)$, yielding
	\begin{equation}
		\label{inequation:dscapacity1gen1}
		Q(\Phi^{\rm DS}) \leq Q(\Phi_{XX'})\;,
	\end{equation}	 and hence the thesis.
\end{proof}
\end{document}